\documentclass[12pt]{article}
\usepackage[utf8]{inputenc}
\usepackage{geometry}
\usepackage{times}

\usepackage[backend=biber,style=ieee,minbibnames=6,maxbibnames=6,doi=false,isbn=false,url=false,eprint=false]{biblatex}
\defbibheading{bibliography}[\refname]{}

\usepackage[utf8]{inputenc}
\usepackage{amssymb}
\usepackage{float}
\usepackage{textcomp}
\usepackage{stfloats}
\usepackage{url}
\usepackage{verbatim}
\usepackage{graphicx}
\usepackage{amsmath}
\usepackage{comment}
\usepackage{bm}
\usepackage{caption}
\usepackage{calc}
\usepackage{subcaption}
\usepackage{setspace}

\title{Learning Separated Representations for Instrument-based Music Similarity}

\author{Yuka Hashizume \\ Graduate School of Informatics, Nagoya University, Nagoya, Japan 
\and Li Li \\ Information Technology Center, Nagoya University, Aichi, Japan,
\and Atsushi Miyashita \\ Graduate School of Informatics, Nagoya University, Nagoya, Japan,
Tomoki Toda \\ Information Technology Center, Nagoya
University, Nagoya, Japan}

\begin{document}
\maketitle

\begin{abstract}
A flexible recommendation and retrieval system requires music similarity in terms of multiple partial elements of musical pieces to allow users to select the element they want to focus on. A method for music similarity learning using multiple networks with individual instrumental signals is effective but faces the problem that using each clean instrumental signal as a query is impractical for retrieval systems and using separated instrumental signals reduces accuracy owing to artifacts. In this paper, we present instrumental-part-based music similarity learning with a single network that takes mixed signals as input instead of individual instrumental signals. Specifically, we designed a single similarity embedding space with separated subspaces for each instrument, extracted by Conditional Similarity Networks, which are trained using the triplet loss with masks. Experimental results showed that (1) the proposed method can obtain more accurate embedding representation than using individual networks using separated signals as input in the evaluation of an instrument that had low accuracy, (2) each sub-embedding space can hold the characteristics of the corresponding instrument, and (3) the selection of similar musical pieces focusing on each instrumental sound by the proposed method can obtain human acceptance, especially when focusing on timbre.
\end{abstract}

\section{Introduction}
\label{intro}
Today's music market is proliferating, with digital products and online services at its core. In 2023, sales from music streaming services increased by 10.4\% from the previous year to 19.3 billion dollars, and streaming services account for 67.3\% of the music industry's sales share~\cite{ifpi}. 
The number of musical pieces available on music streaming services is about 100 million today~\cite{apple}. 
Subsequently, it is impossible for users to listen to all of them to find their favorite music. Therefore, Music Information Retrieval (MIR) technologies, such as music recommendation systems, are needed to help users find their favorite music efficiently.
\par
Research in the field of MIR involves several tasks, mainly the retrieval, recommendation, and estimation tasks to obtain information useful for the retrieval or recommendation, such as tagging, genre classification, melody extraction, and cover song identification. Among these tasks, our focus in this study is to help users find new favorite musical pieces given the background described above. Within the retrieval task, research aimed at such a purpose is ill-defined because it does not have an objective answer unlike cover song retrieval or version retrieval. Hence, various approaches are possible depending on the criteria considered to retrieve or recommend music.
\par
The most common way to access music online today is through text-based metadata retrieval. However, metadata retrieval has limitations in expressiveness when using objective information such as the artist name and publication year~\cite{casey2008}.
On the other hand, music recommendation is a typical technique for users to efficiently discover new favorite music. There are several main approaches in music recommendation~\cite{song2012}: an approach using user information~\cite{dror2011, dror2012}, a content-based approach~\cite{logan2004, oord2013, fathollahi2021}, and a combination of the two~\cite{chen2005, yoshii2006, mcfee2012}. Many studies have been conducted using users' listening history, and one of the most representative approaches is using collaborative filtering~\cite{goldberg1992}. In collaborative filtering, it is assumed that users who have similar ratings or the same behavior toward a certain content will have similar ratings toward other contents. This enables the prediction of the ratings of unknown tracks on the basis of the ratings of other users who have engaged in similar behavior. However, a limitation is that newly released music may receive few recommendations until a certain amount of listening history has been recorded. Another problem is that less well-known music may not be recommended as often since popular music generally receives more ratings.
\par
The content-based approach has the potential to avoid the problems of metadata-based retrieval and collaborative filtering-based recommendation, which uses the features of the content itself for recommendation and retrieval. Content-based methods for recommendation, retrieval, and related MIR tasks have been investigated for a long time~\cite{casey2008} and have traditionally included signal processing, handcrafted features, and classical machine learning methods~\cite{fujishima1999, tzane2001,logan2001, whitman2002, gomez2006}. Fujishima~\cite{fujishima1999} applied pattern matching to acoustic features, Logan and Salomon~\cite{logan2001} applied clustering to acoustic features, and Tzanetakis and Cook~\cite{tzane2001} introduced several handcrafted features for genre classification. Whitman and Rifkin~\cite{whitman2002} proposed the query-by-description method using a classical machine learning method, and Gómez~\cite{gomez2006} presented a method to extract a tonal description from audio signals.
With the advent of deep learning, data-driven embedding extraction has been shown to be effective in improving the performance of MIR systems~\cite{hamel2010, elbir2020, iliadis2022, choi2018}. Hamel and Eck~\cite{hamel2010} showed that using the embeddings extracted from Deep Belief Networks is better than using the Mel-Frequency Cepstrum Coefficient (MFCC) in the classification tasks, and Elbir and Aydin~\cite{elbir2020} also showed that deep learning methods outperform classical machine learning methods in accuracy for genre classification tasks.
Furthermore, the effectiveness of using Convolutional Neural Networks (CNNs) was shown~\cite{oord2013, li2015, hirvonen2015, choi2015, choi2016, lostanlen2016, costa2017, senac2017, park2018, choi2019, pretet2020, wise2021}. These studies include a method for a recommendation system~\cite{oord2013}, an automatic musical instrument identification method~\cite{li2015, lostanlen2016,wise2021}, an autotagging method~\cite{choi2016,choi2019}, a music classification method~\cite{hirvonen2015, costa2017,senac2017,choi2019}, a representation learning method~\cite{park2018, pretet2020}, and analysis of mechanisms~\cite{choi2015}. 
\par
One effective method for content-based music recommendation or retrieval is to define similarities between musical pieces and use the user's favorite piece as a query for retrieval or recommendation. This method requires designing suitable similarity criteria for calculating music similarity, and many MIR researchers discussed this issue~\cite{fathollahi2021, mcfee2012, logan2001, park2018,pretet2020, aucouturier2002, pampalk2006, casey2006, urbano2011,  hamel2013, schlter2013, wolff2014, wolff2015, lu2017, clevelan2020, cheng2020, gabbolini2021}. Aucouturier and Pachet~\cite{aucouturier2002} introduced similarity measures based on a Gaussian model of the cepstrum coefficient. Cheng et al.~\cite{cheng2020} showed some acoustic features related to human perceptions. Moreover, several methods based on machine learning and algorithmic approaches were proposed: a method based on classification~\cite{pampalk2006}, string matching~\cite{casey2006}, learning binary codes for music representations~\cite{schlter2013}, and a path-based music similarity measure~\cite{gabbolini2021}. Urbano et al.~\cite{urbano2011} analyzed the reliability of the results in the evaluation of music similarity.
Some data-driven methods for calculating similarity were proposed: using a classification model~\cite{fathollahi2021, hamel2013}, the metric learning~\cite{mcfee2012, wolff2014, clevelan2020}, and transfer learning~\cite{hamel2013, wolff2015}. McFee et al. proposed a training method with sampling using collaborative filter data~\cite{mcfee2012}. Furthermore, some methods were proposed to learn embedding representation by deep metric learning using labels or tags~\cite{park2018, lu2017, pretet2020}.
In these methods, music similarity is calculated by evaluating a mixture of various sounds using a single criterion. However, music has a complex structure with various significant elements, and what users focus on when listening to music varies from user to user. The MIR system with music similarity regarding multiple partial elements in musical pieces enables users to select the element they want to focus on and flexibly search for music.
\par
We previously proposed a music similarity learning method based on each instrumental part, where networks are trained for each instrument using single instrumental signals with deep metric learning~\cite{has2022}. One limitation of this method is the need for individual instrumental signals not only in training but also in inference, where clean instrumental signals in the query piece the users want to input are usually not publicly available. We also investigated the use of instrumental signals separated from mixed signals, but using the separated signals resulted in lower accuracy than using the clean instrumental signals owing to artifacts.
\par
Another potential approach is to extract an embedding representation for each instrumental part directly from the musical piece.
Several methods have been proposed to extract several different conceptual representations from a single input~\cite{bengio2013}, for example, to disentangle the speaker identity and noise in the speech domain~\cite{hsu2018, hsu2019}, timbre and pitch information in the music domain~\cite{hung2018, luo2019, tanaka21}, and so on.
Veit et al.~\cite{veit2017} proposed Conditional Similarity Networks (CSNs) in the image domain, which learn embeddings differentiated into semantically distinct subspaces that capture the different notions of similarities. Lee et al.~\cite{lee2020} applied CSNs to the music domain and designed an embedding space such that each subspace represents the four similarity metrics: genre, mood, instrumentation, and tempo.
\par
\begin{figure}[t]
\centering
\scalebox{0.3}{
\centerline{\includegraphics{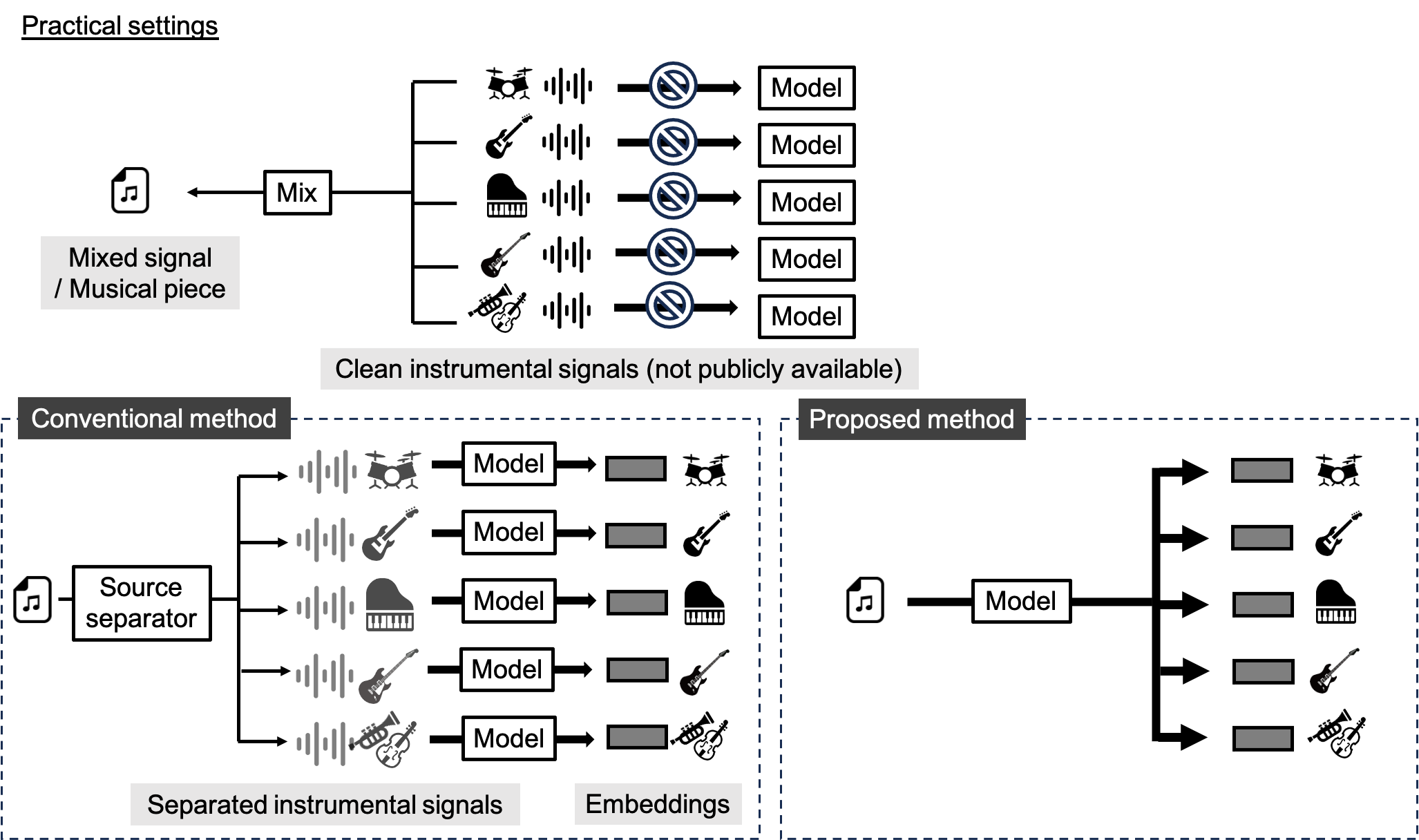}}}
\caption{An overview of limitations in practical settings and the respective approaches of the conventional method and the proposed method. The individual instrumental signals used to create a musical piece (referred to as clean instrumental signals) are not available publicly in general. In the previous study, estimated individual instrumental signals using the source separation model (referred to as separated instrumental signals) were used as input instead of clean instrumental signals, and individual embedding models were trained for each. In this paper, we propose a method for extracting embedding representations based on individual instruments by inputting mixed signals (equal to musical pieces).}
\label{pipeline}
\end{figure}

In this paper, we propose a method for calculating an instrumental-part-based music similarity in one network using musical pieces as input by employing CSNs. The proposed network is trained with deep metric learning to embed musical pieces into a differentiated embedding space, where each subspace selected by a binary mask represents a musical feature when focusing on a particular instrumental part. To successfully train the network, we implement new ideas for the training, such as the use of pseudo musical pieces, a norm loss, and pre-training.
In the experiments, we investigate whether more accurate embedding representations can be obtained using our proposed method than using conventional methods, whether each subspace holds the characteristics of the assigned instrument sounds, and whether the learned similarity criterion matches human perception. The position of the proposed method is shown in Figure~\ref{pipeline}.

\section{Related Work}
\label{relate}
\subsection{Instrumental-part-based similarity with individual networks}
\label{individual}
In deep metric learning with a triplet loss~\cite{ailon2015}, a distance metric is trained with a triplet of samples, where one is considered as an anchor, and the other two are considered as positive and negative samples. Here, the positive sample should have a higher similarity to the anchor than the negative one does. Lee et al.~\cite{lee2020} proposed an unsupervised learning method for music similarity using track information; temporal segments of the same musical track as the anchor are defined as positive samples, and those of different musical tracks from the anchor are defined as negative samples.
\par
To achieve a highly flexible MIR system, we proposed a music similarity learning method that focuses on each instrumental part~\cite{has2022}. In this method, metric learning with triplet loss is applied to individual instrumental signals such as drums, bass, piano, and guitar. We introduced an unsupervised learning approach similar to~\cite{lee2020}, where positive and negative samples are sampled using track information as a substitute for similarity labels for individual instruments. Different networks are separately trained for individual instrumental sounds. This method requires individual instrumental signals not only in training but also in inference, although clean instrumental signals in the query musical piece the users want to input are usually not publicly available. Thus, we applied instrument signals separated from the mixed signals through a music source separation method~\cite{hennequin2020} to this method for evaluation in practical use.
\par
Letting $x_i^{(\rm{a})}$, $x_i^{(\rm{p})}$, and $x_i^{(\rm{n})}$ denote embedding representaions of $i$-th anchor, positive sample, and negative sample, respectively, we constructed the $i$-th triplet as a set of $\{x_i^{(\rm{a})}, x_i^{(\rm{p})}, x_i^{(\rm{n})}\}$, where $i = 1, \ldots, I$ denotes the index of training samples. The triplet loss is defined as
\begin{align}
    &\mathcal{L}_{\rm{triplet}}(x^{(\rm{a})}_i,x^{(\rm{p})}_i, x^{(\rm{n})}_i)\notag\\
    &=\max\{d(x^{(\rm{a})}_i,x^{(\rm{p})}_i)-d(x^{(\rm{a})}_i,x^{(\rm{n})}_i)+\delta,0\},
\end{align}
where $d$ is a distance function for measuring the distance between two embedding representaions, such as the Euclidean distance, and $\delta$ is a margin value, which defines the minimum distance between the positive and negative samples.

\subsection{CSNs}
\label{CSN}
To measure the similarity between images considering multiple notions of similarity, Veit et al. \cite{veit2017} proposed CSNs that learn embeddings differentiated into semantically distinct subspaces that capture the different notions of similarities. In the example where the input is an image of a shoe, the notions of similarity are, for example, the height of the shoes’ heels and the suggested gender of the shoes.
\par
In this method, a network extracting an embedding representation is trained by the triplet loss using masks. For the triplet loss, samples $x^{(\rm{a})}$, $x^{(\rm{p})}$, and $x^{(\rm{n})}$ are selected according to condition $c$ that is defined as a certain notion of similarity. Namely, in the notion corresponding to condition $c$, $x^{(\rm{p})}$ is more like $x^{(\rm{a})}$ than $x^{(\rm{n})}$.
To differentiate the embedding space, a mask is applied to all dimensions except the dimension corresponding to the notion to be considered in the triplet loss calculation. The network is given by function $f(\cdot)$, and $\mathbf{m}_c$ is a mask that activates only the dimension corresponding to condition $c$. The masked distance function between two images $x_i$ and $x_j$ is given by
\begin {align}
\label{CSNtriplet}
d(x_i,x_j;\mathbf{m}_c)=\parallel f(x_i)\mathbf{m}_c-f(x_j)\mathbf{m}_c\parallel_2.
\end{align}
Thus, the triplet loss can be written as
\begin{align}
 &\mathcal{L}_{\rm{triplet}}(x^{(\rm{a})},x^{(\rm{p})},x^{(\rm{n})},c)\notag \\
 &=\max\{d(x^{(\rm{a})},x^{(\rm{p})};\mathbf{m}_c)-d(x^{(\rm{a})},x^{(\rm{n})};\mathbf{m}_c)+\delta, 0\}.
\end{align}
\par
Lee et al. \cite{lee2020} proposed the disentangled multidimensional metric learning for music similarity using CSNs. They used musical genre, mood, instrument, and tempo for the notions of similarity. 

\begin{figure}[t]
\centering
\scalebox{0.45}{
\centerline{\includegraphics{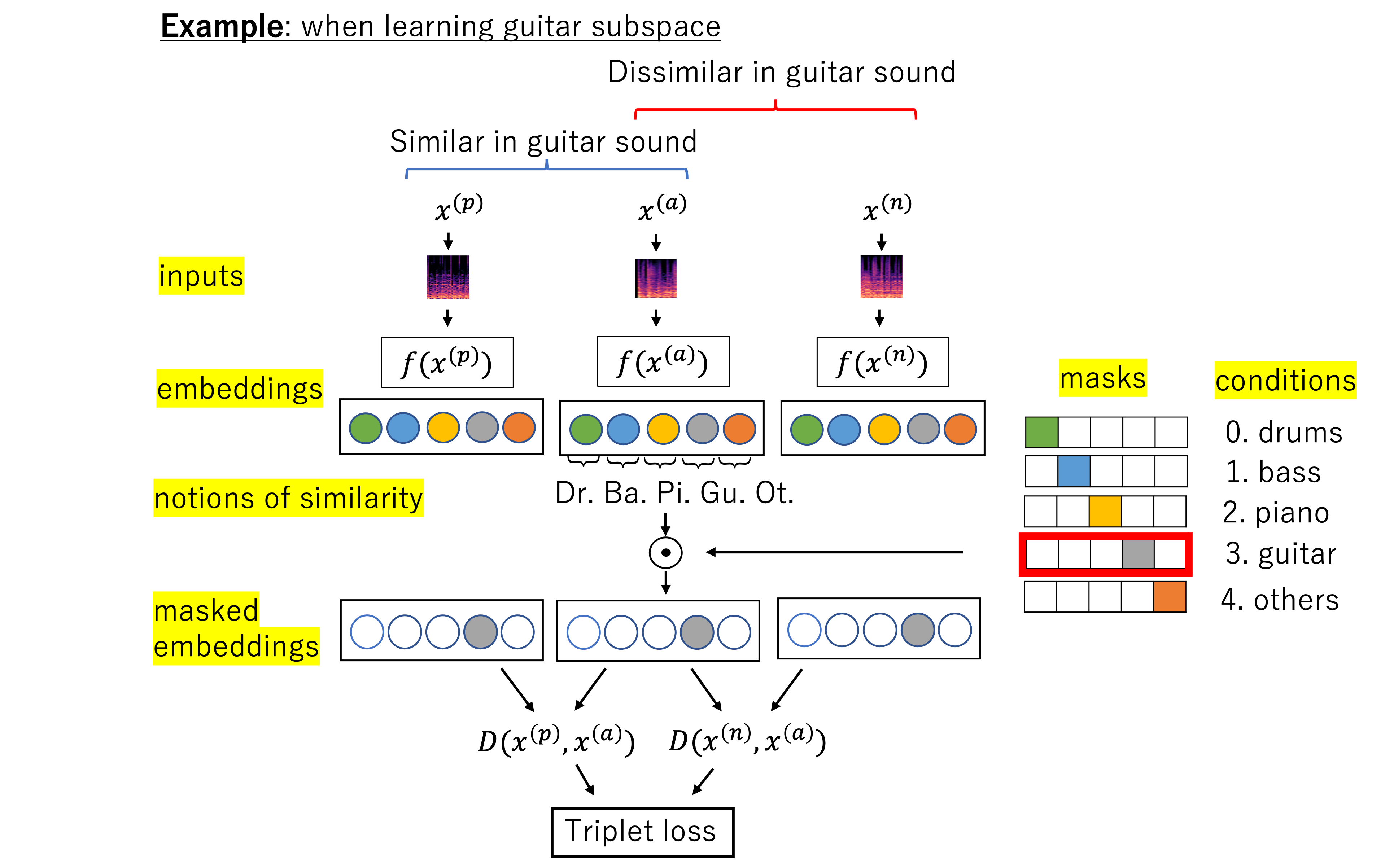}}}
\caption{Overview of the proposed method. $x^{(\rm{a})}$, $x^{(\rm{p})}$, and  $x^{(\rm{n})}$ denote the anchor, positive, and negative samples, respectively. “Dr.,” “Ba.,” “Pi.,” “Gu.,” and “Ot.” are drums, bass, piano, guitar, and others, respectively. This figure shows an example of setting the condition to $c=3$, i.e., similarity focusing on the guitar part, where an anchor sample $x^{(\rm{a})}$ and a positive sample $x^{(\rm{p})}$ are similar, and an anchor sample $x^{(\rm{a})}$ and a negative sample $x^{(\rm{n})}$ are dissimilar when focusing on the guitar part. From each sample, the embedded representation is extracted by the network and is masked so that the subspace to which the guitar is assigned only validates in the triplet loss calculation.}
\label{prop}
\end{figure}

\section{Proposed method}
\label{prop-method}
\subsection{Triplet loss with mask}
\label{triplet}
In this study, the CSNs described in Section~\ref{CSN} are used, with each notion of similarity defined as each instrumental-part-based similarity. We define $c$ as the condition where $c=0, 1, 2, 3, 4$ represent the similarity based on drums, bass, piano, guitar, and others, respectively, to differentiate the embedding space into subspaces that each represent similarities based on individual instrumental sounds. Letting $D$ be the number of dimensions of a subspace assigned for one instrumental part, the subspace of the embedding assigned to condition $c$ is $f(x)[cD:(c+1)D-1]$, where $f(x)$ is an output of the network.
The following formula defines each element of a $(5\times D)$-dimensional vector $\mathbf{m}_c$ as a mask that keeps the subspace corresponding to $c$ and sets the other dimensions to 0, with $k$ being the dimension index.
\begin {align}
m_{ck} = \left\{
\begin{array}{ll}
1, & (cD\leqq k<(c+1)D)\\
0, & (\mbox{otherwise}).
\end{array}
\right.
\end{align}
The triplet loss in CSNs shown in Equation~\ref{CSNtriplet} is used for training with the mask described above. An overview of the proposed method is shown in Figure~\ref{prop}.

\subsection{Norm loss}
\label{norm}
In our method, it is required to prevent any leakage of features of the instrumental parts to the unassigned subspaces. When the input music does not contain some instrumental parts, we add the constraint to output a value close to a zero vector in the subspace corresponding to those instrumental parts. This constraint at least ensures that if the input does not contain some instrumental parts, the corresponding subspaces do not contain any values calculated from other instrumental parts' signals. 
\par
\begin{figure}[t]
\centering
\scalebox{0.48}{
\centerline{\includegraphics{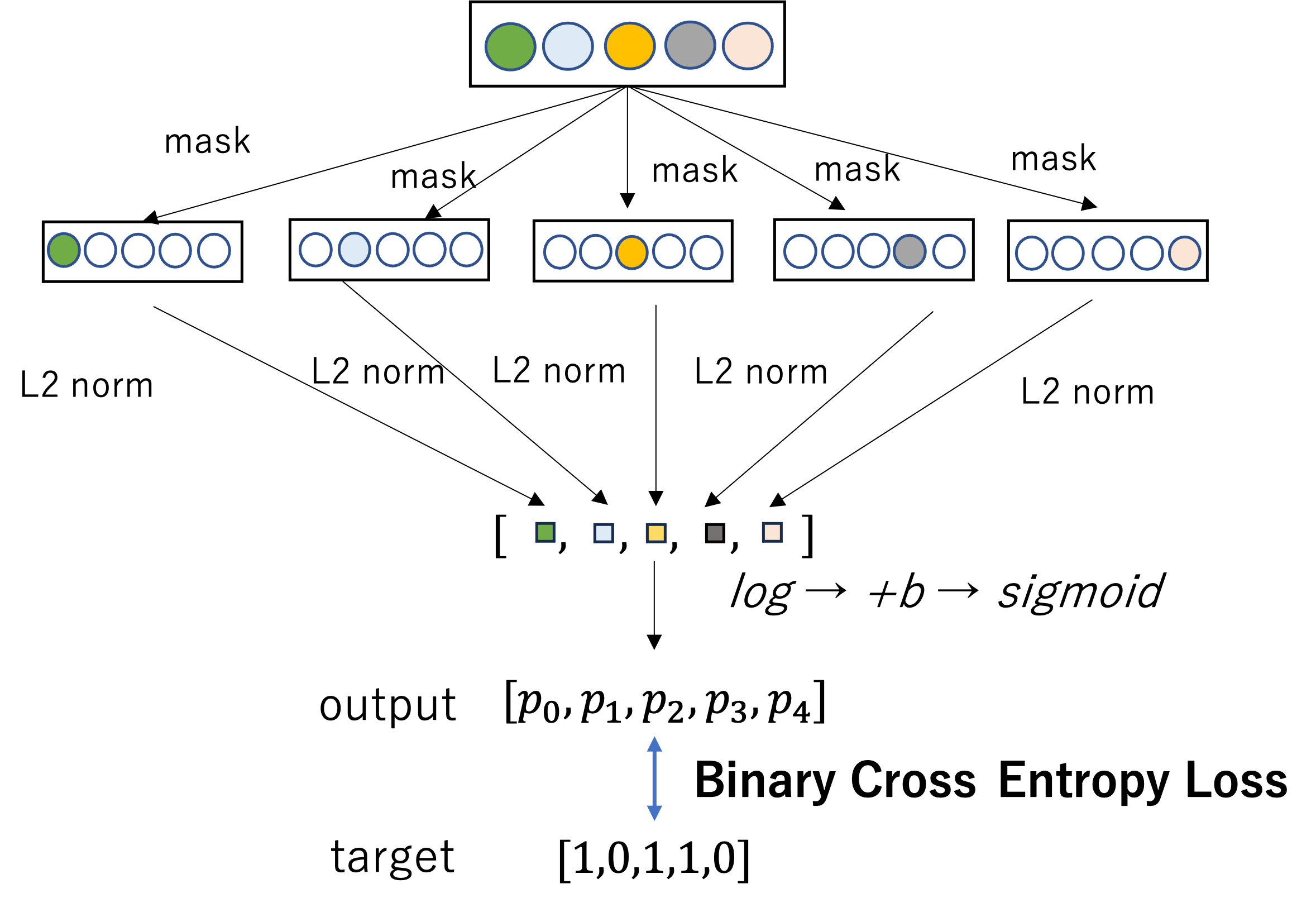}}}
\caption{Procedures for calculating norm loss. This is an example of when a sound containing only drums, piano, and guitar is input, where the bass's subspace and other's subspace are trained to be close to the zero vector.}
\label{normloss}
\end{figure}

We use the Binary Cross Entropy Loss (BCELoss) to satisfy this constraint. The input of the BCELoss is a five-dimensional vector $\mathbf{p}_i$, whose values are calculated from the norm of each subspace for $x_i$. Each value of $p_{ij}, (j=0,1,2,3,4)$ is calculated by taking the logarithm to the norm of the masked embedding $f(x_i)\mathbf{m}_c, (c=j)$ and then adding a learnable parameter $b_i$: 
\begin{align}
&p_{ij}=\sigma( \log( ||f(x_i)\mathbf{m}_j||_2 )+b_{ij}),
\end{align}
where $\sigma(x) = \frac{1}{1 + e^{-x}}.$ The target is a five-dimensional multi-hot vector $\mathbf{q_i}$ that is set to 1 if each instrumental sound is included in the input and 0 if not: 
\begin {align}
&q_{ij} = \left\{
\begin{array}{ll}
1, & (P_{\rm{avg}}(x_{ij})>threshold)\notag\\
0, & (\mbox{otherwise}),
\end{array}
\right.\\
\end{align}
where $P_{\rm{avg}}$ means a time average power of the signal, and $x_{ij}$ is a clean individual instrumental signal contained in $x_i$, where the subscript represents each instrument. When $\mathbf{p}_i$ computed from the $i$-th anchor $x^{(\rm{a})}_i$ is denoted as $\mathbf{p}^{(\rm{a})}_i$, and in the same way for a positive sample and a negative sample, the formulation of the norm loss is as follows.
\begin {align}
\label{bce}
&\mathcal{L}_{\rm{norm}}(x^{(\rm{a})}_i,x^{(\rm{p})}_i, x^{(\rm{n})}_i)\notag\\
&=\frac{1}{3}\{BCE(\mathbf{p}^{(\rm{a})}_i, \mathbf{q}^{(\rm{a})}_i)\notag\\
&+BCE(\mathbf{p}^{(\rm{p})}_i, \mathbf{q}^{(\rm{p})}_i)\notag\\
&+BCE(\mathbf{p}^{(\rm{n})}_i, \mathbf{q}^{(\rm{n})}_i)\},\notag\\
&BCE(\mathbf{p}, \mathbf{q})=\frac{1}{5}\sum_{j=0}^{4}\left\{{q_j \log p_j + (1-q_j) \log(1-p_j)}\right\}.
\end{align}
This procedure is shown in Figure \ref{normloss}. Note that we use not only the observed silent sections for each instrumental part but also the created silent sections, i.e., we arbitrarily mix only some instruments from the instruments contained in a musical piece to equalize the total time of each instrumental sound included in the train data.

\subsection{Training network}
\label{training}
The final loss function $\mathcal{L}$ is as follows, where $\lambda$ is the hyperparameter that weights the loss function $\mathcal{L}_{\rm{norm}}$. Note that this loss function is averaged within the mini-batch.
\begin {align}
\mathcal{L}=\mathcal{L}_{\rm{triplet}}+\lambda \mathcal{L}_{\rm{norm}}
\end{align}

\subsection{Pseudo musical piece}
\label{pseudo}
As mentioned in Section~\ref{triplet}, we need to sample triplets for each notion of similarity, i.e., a similar/dissimilar pair focusing on each instrumental part. However, there is no label that evaluates whether musical pieces are similar to each other focusing on each instrumental part. Therefore, we use the unsupervised learning method similar to the previous methods. However, using the track information as a label, as illustrated in Figure~\ref{nopsdtrip}, is not appropriate to obtain distinct representations for each notion of similarity. This is because the labels used for training the respective subspaces are the same over all subspaces for a single input, even though the goal is to represent multiple distinct representations from one input.
\par
To successfully train the separated subspaces, we propose a method to create a pseudo musical piece for input by mixing instrumental signals in different musical pieces. For example, when the drum sound contained in piece A is called drum sound A, a pseudo musical piece can be created by mixing the drum sound A with other instrumental sounds from another piece B. We denote this piece's label as $\rm{(A, B)^{(dr, else)}}$ and also call this piece with the drums label A. By this method, we can create a pair such as one that has the same drum label but different guitar labels. To distinguish the pseudo musical pieces, we refer to the original musical pieces included in the dataset as \textit{dataset musical pieces} throughout this paper.
\subsubsection{Basic triplet}
Considering that musical pieces with the same label for a particular instrument are similar to each other on that instrument, a triplet sample can be created. We can say that segment 1, randomly extracted from the musical piece with label $\rm{(A, B)^{(dr, else)}}$, and segment 2, randomly extracted from the musical piece with label $\rm{(A, C)^{(dr, else)}}$, are similar in drum sounds but dissimilar in sounds other than drums. On the other hand, segment 1 and segment 3 with label $\rm{(D, B)^{(dr, else)}}$ are dissimilar in drum sounds but similar in other sounds. Therefore, segments with label \{$\rm{(A, B)^{(dr, else)}}$, $\rm{(A, C)^{(dr, else)}}$, $\rm{(D, B)^{(dr, else)}}$\} can be used as an anchor, a positive sample, and a negative sample in learning with condition $c=0$. We can also sample the triplets for other conditions $c=1,2,3,4$ in the same way. The triplets extracted in this way are called the basic triplets.
\subsubsection{Additional triplet}
We further add triplets of interchanged positive and negative samples under a condition other than that of the basic triplet to allow each subspace to explicitly learn a different similarity criterion. The anchor and the negative sample in the above example, the segments from pieces with labels $\rm{(A, B)^{(dr, else)}}$ and $\rm{(D, B)^{(dr, else)}}$, are dissimilar in drum sounds but similar in instrumental sounds other than drum sounds. Thus, samples with label \{$\rm{(A, B)^{(dr, else)}}$, $\rm{(D, B)^{(dr, else)}}$, $\rm{(A, C)^{(dr, else)}}$\} can be used as an anchor, a positive sample, and a negative sample in learning with condition $c\neq0$. 
By randomly selecting $c$ from those that are different from the basic triplet, an additional triplet is created for each basic triplet.
\par
\begin{figure}[t]
\centering
\scalebox{0.4}{
\centerline{\includegraphics{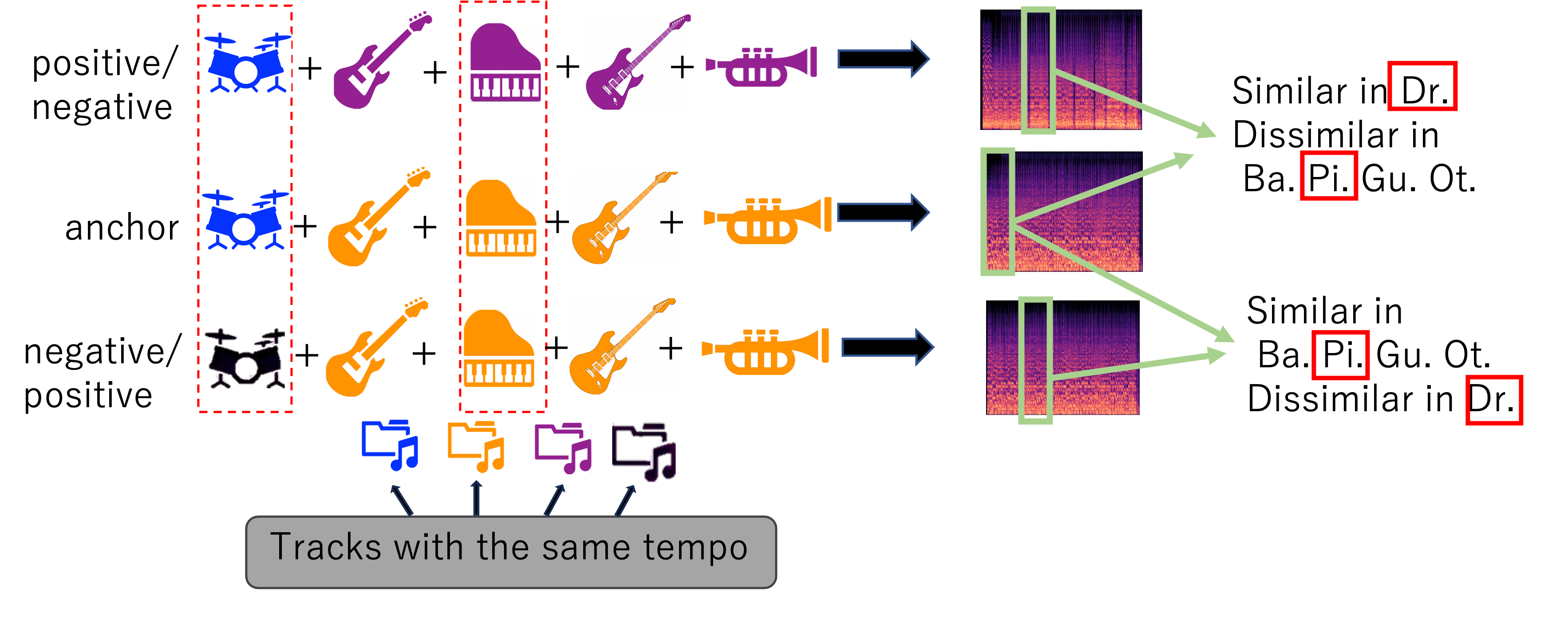}}}
\caption{How to create the basic triplet and the additional triplet with pseudo musical pieces. Each color represents an ID in the dataset musical piece, and four musical pieces are randomly selected from the same tempo class in the train set, and pseudo musical pieces are created using the individual instrument signals in them. When these triplet samples are input, two losses are calculated: a loss where the upper sample is calculated to be close to the anchor in the drum space, and a loss where the lower sample is calculated to be close to the anchor in the piano space.}
\label{trip2}
\end{figure}

An example of these triplet extraction processes is shown in Figure~\ref{trip2}. Note that negative samples in the basic triplet are selected in such a way that additional triplets can be constructed. Specifically, in this example, the instrumental sounds other than drums share the same labels as those of the anchor.

\begin{figure}[t]
\centering
\scalebox{0.45}{
\centerline{\includegraphics{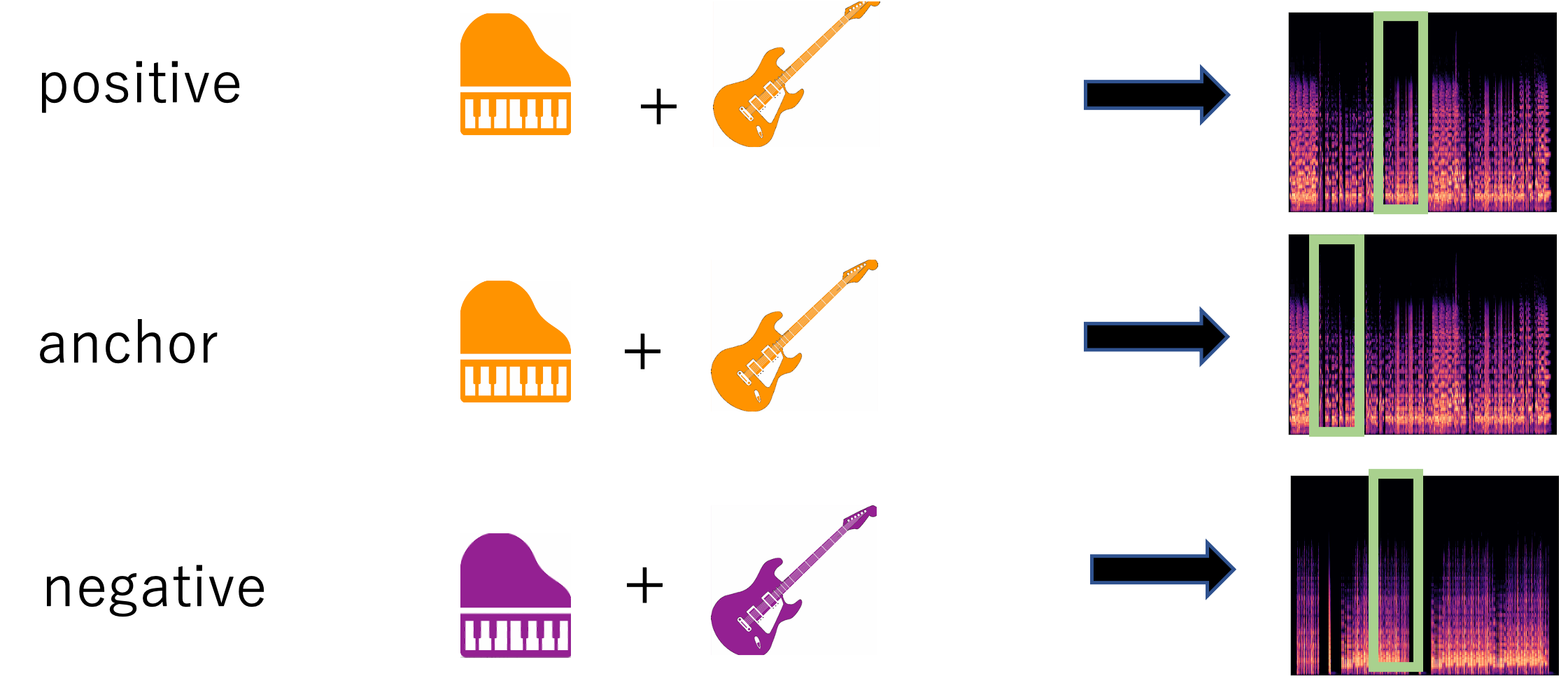}}}
\caption{How to create a triplet using track information with the dataset musical pieces instead of pseudo musical pieces. We show the example of combining only some instruments for the calculation of the norm loss.}
\label{nopsdtrip}
\end{figure}

\subsection{Pre-training}
\begin{figure}[t]
\centering
\scalebox{0.5}{
\centerline{\includegraphics{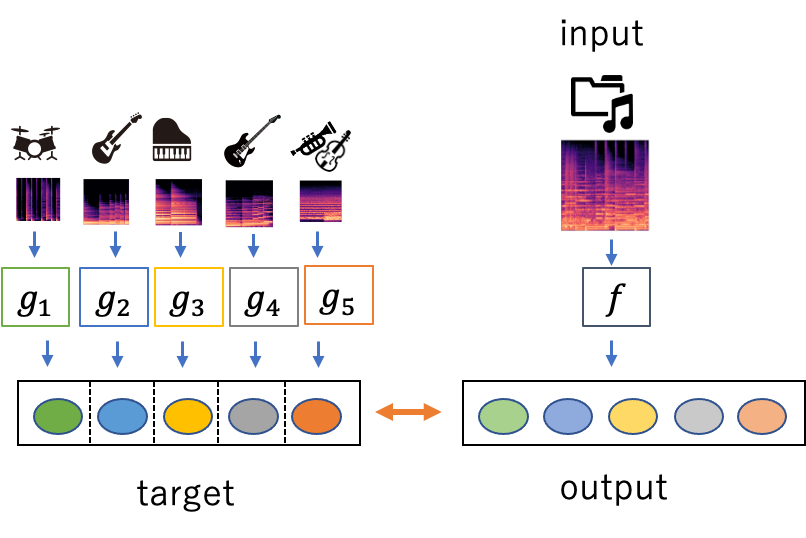}}}
\caption{Procedures for pre-training. The network $f$ is trained so that its output is close to the target by the MSE loss. The input of the network is the segments of the dataset musical pieces. The target is the concatenation of embeddings extracted from the individual networks.}
\label{pretrain}
\end{figure}

We introduce pre-training to start training from a better initial value than a random value, enabling the above learning methods to be effective. We train the network with the Mean Squared Error Loss (MSELoss) using the output of the individual instrumental-part-based similarity network~\cite{has2022} as the ground truth for each subspace. Namely, the concatenation of $g_j(x_{ij})$ is the target of pre-training, where $g_j( \cdot),  (j=0,1,2,3,4)$ are denoted as the individual networks corresponding to drums, bass, piano, guitar, and others. For the same reasons explained in Section~\ref{norm}, the ground truth sub-embedding is set to the zero vector if the input musical piece does not contain the corresponding instrumental sound. Figure~\ref{pretrain} shows a way of creating the target embeddings. $x_{ij}$ is the clean instrumental segment of instrument $j$ contained in the $i$-th the dataset musical piece's segment $x_i$. $\frac{\mathbf{y}_i}{\parallel \mathbf{y}_i \parallel_2}$ is the target embedding for the network training, which is created by concatenating embeddings extracted from $x_{ij}, (j=0,1,2,3,4)$ using the individual networks and divided by the norm. The formulations of the loss function of pre-training $\mathcal{L}_{\rm{pre}}$ and the target embedding are as follows:

\begin{align}
&\mathcal{L}_{\rm{pre}}(x_i)= \left(f(x_i) - \frac{\mathbf{y}_i}{\parallel \mathbf{y}_i \parallel_2}\right)^2 ,\notag
\\&y_{ik} = g_j(x_{ij})_k,  (jD\leqq k<(j+1)D).
\end{align}

This loss function is averaged within the mini-batch.

\begin{figure}[t]
\centering
\scalebox{0.33}{
\centerline{\includegraphics{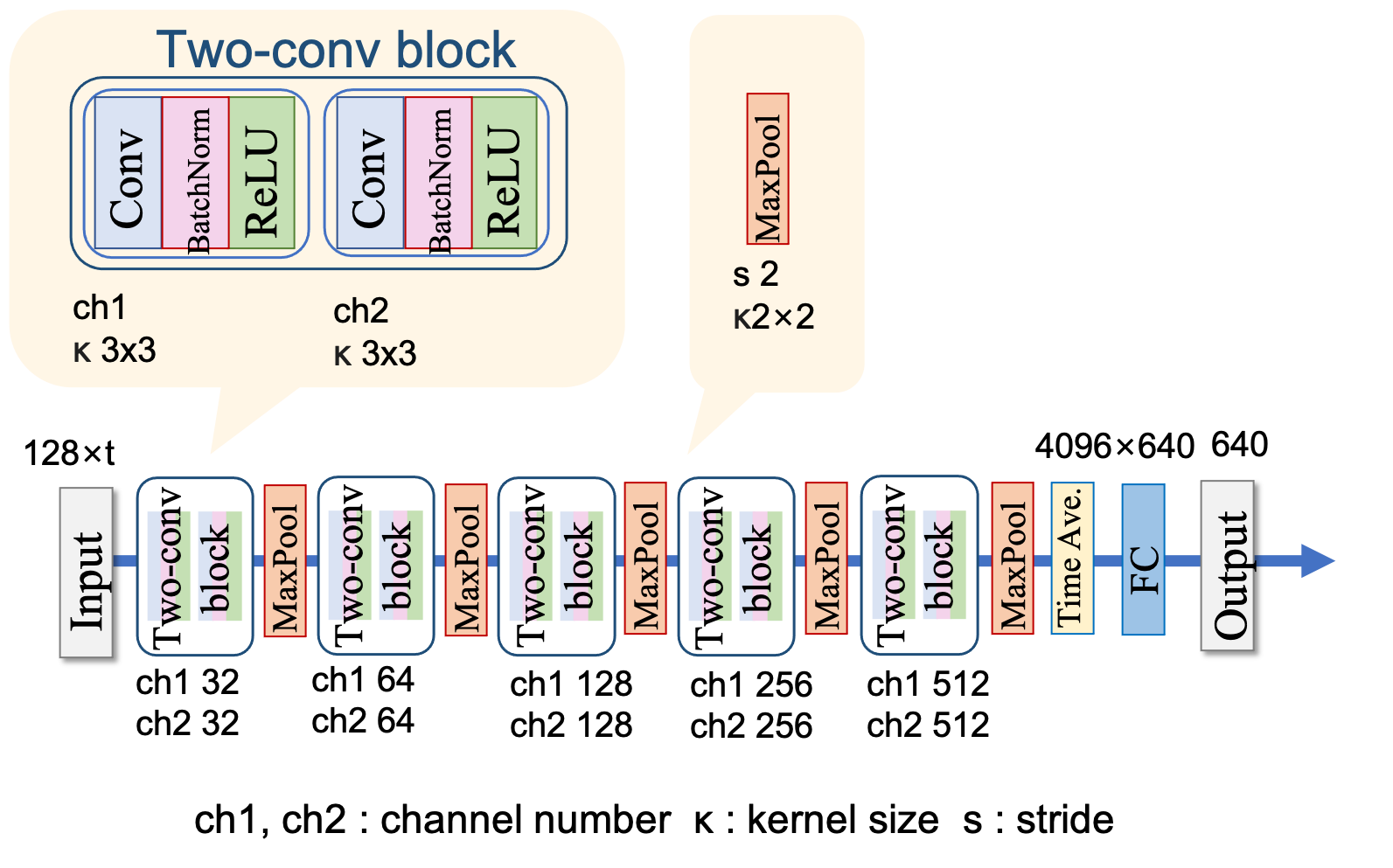}}}
\caption{Network architecture. “ch1” and “ch2” denote the channel number, and “$\kappa$” and “s” denote kernel size and stride, respectively. “Conv” and “FC” denote the convolutional and fully connected layers, respectively. “Time Ave.” means to take an average in the time direction. The numbers above input, output, and “FC” are their sizes. “t” is calculated by multiplying the number of seconds by the sampling rate and dividing it by the hop length.}
\label{network}
\end{figure}

\section{Experimental evaluation}
\subsection{Experimental conditions}
\subsubsection{Dataset and input features}
The dataset we used is Slakh2100 \cite{manilow2019}, which contains non-vocal musical pieces and their stems. Following Slakh's recipe, individual instrumental sounds, namely, drums, bass, piano, and guitar sounds, were created from their stems, and the stems that did not fit into any of the four instruments were mixed as “others.” We used the \textit{redux} subset of Slakh2100, which is created by omitting some tracks so that each MIDI file only occurs once. The redux subset has a total of 1710 tracks: 1289 in train, 270 in validation, and 151 in test.
\par
We used the 200 dataset musical pieces and their clean instrumental sounds in the training set in pre-training, both for training the proposed network with $\mathcal{L}_{\rm{pre}}$ and training individual networks~\cite{has2022}. Moreover, the pseudo musical pieces for training were created with the instrumental sounds contained in the 1200 musical pieces in the training set for training with $\mathcal{L}_{\rm{triplet}}$. The 270 musical pieces in the validation set were used to crerate the pseudo musical pieces for validation. The Slakh test set was used for testing, but musical pieces with the shortest non-silent sections in individual instrument sounds were excluded one by one until 10\% of the musical pieces were removed, after which 136 pieces were used.
\par
When creating the pseudo musical pieces, the dataset was classified into 36 classes according to tempo, and the instrumental sounds contained in musical pieces belonging to the same tempo group were allowed to be mixed together to preserve the music-like nature of the music. 
Under this rule, multiple different pseudo musical pieces containing the same instrumental sound were generated. The 5000 triplet pseudo musical pieces were randomly created every epoch using 1200 musical pieces for metric learning with $\mathcal{L}_{\rm{triplet}}$.
\par
Both the dataset musical pieces and pseudo musical pieces were divided into three-second segments for pre-training and training with 50\% overlap, three-second segments without overlap for validation, and 3-, 5-, and 10-second segments without overlap for testing.
All segments were converted to dB-scaled mel-spectrograms with 2048 window length and 512 hop length, normalized, and used as input for the training, validation, and testing.

\subsubsection{Network}
We used the network shown in Figure~\ref{network}, which had 10 convolutional layers with batch normalization and ReLU, and Max pooling applied every two convolutional layers. The encoder portion of U-Net~\cite{ronneberger2015, jansson2017} was referenced. This network was trained to extract a 640-dimensional embedding vector from a mel-spectrogram as embedding representations. The 640-dimensional embedding representation was aimed to have 128-dimensional subspaces assigned to each of the five instruments. The subspaces were assigned to drums, bass, piano, guitar, and others in order of increasing dimensions.

\subsubsection{Pretraining conditions}
For each instrument, a Convolutional Network was trained as an individual network as in the previous study~\cite{has2022}. Then, we pre-trained the network shown in Figure~\ref{network} using the segments of the dataset musical pieces as inputs and the concatenations of outputs of individual networks as targets. We used the clean instrumental sound segments as input for the individual networks both in training and inference to create the target embeddings.

\subsubsection{Training conditions}
The weighting parameter $\lambda$ between two losses $\mathcal{L}_{\rm{triplet}}$ and $\mathcal{L}_{\rm{norm}}$ was set to 0.1. The margin of the triplet loss function, the number of epochs, and the batch size were set to 0.2, 1000, and 32, respectively.  

\subsubsection{Baseline model}
We used our conventional method~\cite{has2022} as the baseline model. In our previous study, we tried using both clean instrumental signals and separated instrumental signals as input, but in order to make the conditions the same as in this study, where mixed signals are used as input, we used the method of using separated instrumental signals as the baseline. Following the previous study, we used the Spleeter~\cite{hennequin2020} for the sound source separation method and used the separated signals as input for both training and inference.

\subsection{Evaluation method}
We conducted experimental evaluations to investigate whether the following purposes of this study were achieved: (P1) to learn an embedding representation in which similar pieces are close and dissimilar pieces are far from each other more accurately than the conventional method, (P2) to output the similarity focusing on each instrumental part in the subspace assigned to each instrument, (P3) to ensure that the constraints imposed to be satisfied during training are also satisfied during inference, and (P4) to learn similarity criteria corresponding to human perception.
\subsubsection{Accuracy of embedding representation}
\label{knn}
In the evaluation on P1, we used the accuracy of music ID prediction with the dataset musical pieces in the same manner as the evaluation in the previous study~\cite{has2022}. This evaluation was based on the assumption that instrumental sounds that consist of different time segments of the same musical piece should be more similar than those of different musical pieces. Note that subjective evaluation experiments later confirmed whether this assumption fits the human senses.
Specifically, we used the K-nearest neighbor (kNN) method to predict the music IDs of the test segments' representations. Let the segment to be predicted be called the target, and the music IDs of all test segments' representations except the target were assumed to be known. We predicted the music ID of the target by a majority vote using the IDs of the top five nearest test segments' representations, and this was done for all musical pieces in the test set.
To evaluate each instrument's representation, we extracted it by using only the corresponding subspace with masking, inputting the dataset musical pieces with our proposed method. For the evaluation of the conventional method, the same musical pieces were inputted into the source separation model~\cite{hennequin2020} according to the previous study~\cite{has2022}, and the separated signals were input into each individual network to extract each instrument's representations.

\subsubsection{Capability to represent separated embeddings}
\label{pseknn}
We evaluated the accuracy of the embedding representation of each subspace in Section~\ref{knn} but did not evaluate whether each subspace is separated by the instrument. The inputs in the evaluation in Section~\ref{knn} had the same label for all instruments because they were the dataset musical pieces. For example, if a piano feature leaked into the drum space, we were still able to predict the correct drum label using that feature.
\par
In the evaluation on P2, the pseudo musical pieces were created using the musical pieces in the test set by the same method as described in Section~\ref{pseudo}. The kNN-based label prediction by the same method as described in Section~\ref{knn} was conducted for the test pseudo musical pieces, where not only the target but also other segments divided from the same pseudo musical piece as the target were removed from the reference. For example, in the evaluation of the drums subspace, the test musical pieces with the same drum label and different other instrument labels were created, as $\{\rm{(A, B)^{(dr, else)}}, \rm{(A, C)^{(dr, else)}}, \rm{(A, D)^{(dr, else)}}, \rm{(A, E)^{(dr, else)}}$, $ \rm{(F, G)^{(dr, else)}}, \\ \rm{(F, H)^{(dr, else)}}...\}$.
The correct drum label of a target, a segment from the musical piece with the label $\rm{(A, B)^{(dr, else)}}$, is “A” but other segments from the musical piece with $\rm{(A, B)^{(dr, else)}}$ cannot be referred. Hence, only when segments from $\rm{(A, C)^{(dr, else)}}$ or $\rm{(A, D)^{(dr, else)}}$ or $ \rm{(A, E)^{(dr, else)}}$ were close to the target (even though they have different labels except for drums), the prediction works well.
If the drum subspace contains the piano's feature, the prediction is affected by that and can be wrong. This is because segments that are similar on the piano to the target but have different drum labels come close to the target. This can detect the leak and correctly evaluate the capability to represent separated embeddings. We created 40 pseudo musical pieces with 10 labels for each instrument; in other words, four different pseudo musical pieces per label, and divided them into segments.

\subsubsection{Instrumental sound identification accuracy}
\label{recog}
To confirm that the training with the norm loss described in Section~\ref{norm} was successful (P3), we evaluated it with the instrumental sound identification task. We performed this to verify that the subspace corresponding to instrumental sounds not included in the input was close to the 0 vector, i.e., that the information for each instrumental sound did not leak into the subspace to which it did not correspond.
\par
When a clean instrumental signal was input, a five-dimensional vector was calculated such that the $j$-th element had the norm of the masked embedding $f(x)\mathbf{m}_{j}$, and the index with the largest value of the vector was denoted as the prediction of the type of instrumental sound for that input. We made the above predictions using the clean instrumental signals of the musical pieces in the test set and calculated the percentage of correct answers for all instruments.

\subsubsection{Correlation between distance matrices}
\label{distmatrix}
In our method, $f(x_{ij})$, the output when inputting the instrumental sound $x_{ij}$, and $f(x_i)\mathbf{m}_{j}$, the masked embedding when inputting a musical piece $x_i$ containing that instrumental sound $x_{ij}$ should represent the same feature.
We confirm whether our model had a correlation between these embeddings focusing on similarity relationships between musical pieces (one of P3). We calculated the two distance matrices between representations of all segments of the test set (the dataset musical pieces) for $f(x_{ij})$ and $f(x_i)\mathbf{m}_{j}$. Then, we flattened them to single vectors and calculated the correlation between them.
\par
In contrast to the above, a very high correlation between different subspaces may have resulted in a meaningless space that expresses the same feature in all of them when inputting the mixed sound. Correlations between two subspaces when inputting the mixed sound were also calculated to verify the difference between similarity criteria in subspaces (one of P3). Each distance matrix was calculated using $f(x_i)\mathbf{m}_{j}$, and the correlations between distance matrix pairs were calculated in the same manner as above.

\subsubsection{Subjective evaluation}
\label{sub}
Subjective evaluation experiments through a listening test~\cite{has25} were conducted to confirm whether each subspace represented a similarity criterion such that the distance between sounds was small enough that humans would perceive them to be similar when listening to each assigned instrumental sound (P4). The procedure of the listening test is as follows.
\par
Subjects were presented with three audio tracks of instrumental sounds, X, A, and B, and listened to all of them. They chose A+ if they perceived A to be more strongly similar to X than B, A$-$ if slightly, B+ if they perceived B to be more strongly similar to X than A, and B$-$ if slightly, on the basis of the following four perspectives: timbre, rhythm, melody, and overall similarity.
\par
In each answer, they can select N/A from up to two perspectives except for \textit{overall}. The following instruction was provided with the subjects: “You can select N/A if A and B are similar/dissimilar to X of equal degree, or the presented instrumental sound has no element corresponding to the perspective; e.g., drums have no melody.”
\par
We prepared two types of sample sets, test 1 and test 2.
For test 1, we randomly selected three different musical tracks $\{\chi_i, \alpha_i, \beta_i\}$ and randomly captured five-second segments from each instrumental sound contained in the three tracks, respectively. Then we obtained one sample set for each instrument, \{X, A, B\}=$\{\chi_{ij}$'s segment, $\alpha_{ij}$'s segmant, $\beta_{ij}$'s segment\}. The subscript $j$ represents each instrument ($j=0,...,4$), with $0$ representing drums, $1$ bass, $2$ piano, $3$ guitar, and $4$ others. For example, $\chi_i^{(0)}$ means the drum sound contained in the musical piece $\chi_i$. This selection was repeated four times $(i=0,...,3)$, and 20 sample sets were created. For test 2, we used the same X as in test 1, and one of the other two samples was taken from a different time frame of the same song as X. Namely, we randomly selected $\gamma_i$ from the test music tracks excluding $\chi_i$ and replaced $\alpha_i$, $\beta_i$ with $\chi_i$ and $\gamma_i$ (in no particular order). The process was repeated in the same way, and 20 sample sets $\{\chi_{ij}$'s segment 1, $\chi_{ij}$'s segment 2, $\gamma_{ij}$'s segment\} were created. It was randomly determined whether this set was \{X, A, B\} or \{X, B, A\}. Thus, a total of 40 sample sets were created, and these were used as one listening test set.
\par
This procedure was repeated with a random selection of sample
sets, creating a total of 60 listening test sets. Experiments were conducted by recruiting participants through
CrowdWorks~\cite{cw}. The valid answers from a total of 632 subjects (281 unique subjects) were obtained, and for all 60 sets, valid answers from at least six different subjects were obtained.
\par
We calculated whether A or B was closer to X using our proposed model for the same set as that used in the listening test and then calculated the matching rate between the model's results and the subjects' results. The model's results were obtained as follows: The music tracks originally containing the instrumental tracks (A, B, and X) used in the listening test were input into the model, and then the distance was measured by applying a mask that leaves only the subspace corresponding to the target instrument. Then, the result was the one with the smaller distance from A or B to X. On the other hand, the subjects' results were obtained as follows: A+ and A$-$ were treated as the same answer, A. The same applied to B. Sample sets with less than 80\% agreement among subjects were eliminated in the evaluation because the sample set with a low agreement rate among the subjects may be equally similar/dissimilar to X for both A and B. If N/A accounts for the largest percentage (even over 80\% agreement), that sample set was also eliminated. All answers to the remaining sample set, A or B, were used as the subjects' results. 

\subsection{Results}
\subsubsection{Accuracy of embedding}
The accuracy of the predicted music IDs is shown in Table \ref{table:knn}. The table shows the results when 3, 5, and 10 s of data were used as input for inference.
Each row represents the instrument to focus on. The column for the proposed method shows the results of inference using only the subspace to which the focused instrument sound is assigned. In contrast, the column for the baseline shows the results of inputting the separated instrument signals to the individual networks. 
\par
\begin{figure}[ht]
    \begin{minipage}[t]{0.45\columnwidth}
        \centering
        \includegraphics[scale=0.31]{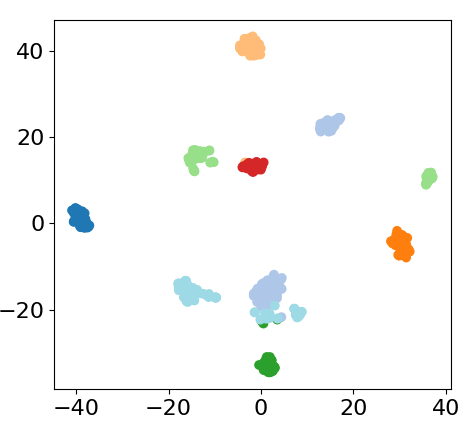}
        \subcaption{Drums subspace}
    \end{minipage}
    \begin{minipage}[t]{0.45\columnwidth}
        \centering
        \includegraphics[scale=0.31]{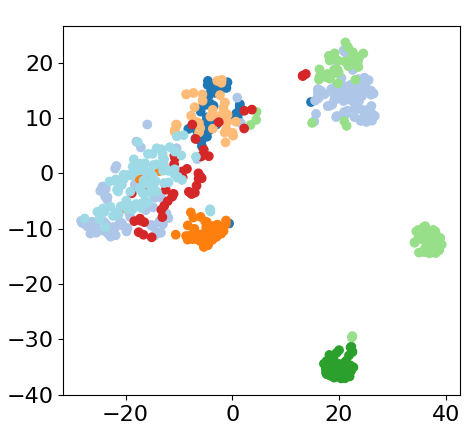}
        \subcaption{Bass subspace}
    \end{minipage}
    \begin{minipage}[t]{0.45\columnwidth}
        \centering
        \includegraphics[scale=0.31]{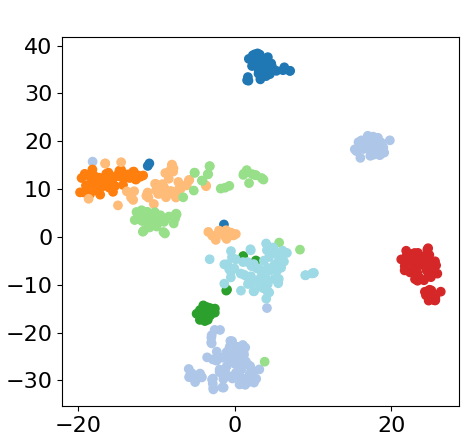}
        \subcaption{Piano subspace}
    \end{minipage}
    \begin{minipage}[t]{0.45\linewidth}
        \centering
        \includegraphics[scale=0.31]{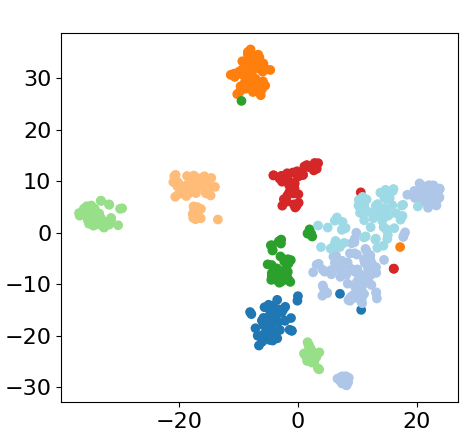}
        \subcaption{Guitar subspace}
    \end{minipage}
    \begin{minipage}[t]{0.45\columnwidth}
        \centering
        \includegraphics[scale=0.31]{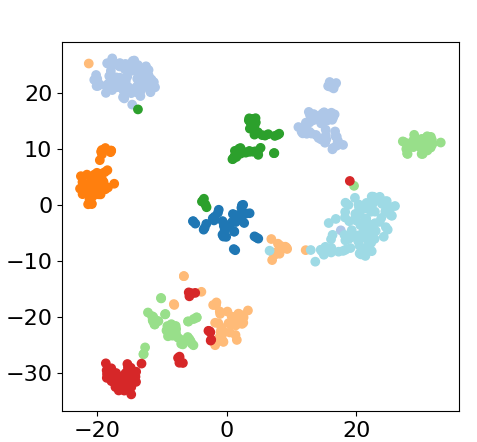}
        \subcaption{Others subspace}
    \end{minipage}
\caption{Examples of visualized embedding representations extracted from five-second segments in the test set of the dataset musical pieces. We used t-SNE \cite{laurens2008} to compress the 640-dimensional embedding representations to two-dimensional representations. Different time frames from the same musical piece are plotted with the same color.}
\label{emb}
\end{figure}

It can be seen that the baseline using separated instrumental signals is affected by sound quality degradation due to separation, and the accuracy of embedding representation degrades, especially on piano with low separation accuracy~\cite{has2022}. In contrast, the proposed method shows stable accuracy regardless of which instrument is focused on.
An example of the subspace is shown in Figure~\ref{emb}. It can be seen that segments from the same musical piece constitute a cluster and that the subspace can be learned with different distance relationships between the musical pieces.

\begin{table}[th]
\centering
\vspace*{1mm}
\scalebox{0.7}{
\begin{tabular}{c|c|c|c|c|c|c}
\hline
&\multicolumn{2}{|c|}{Input: 3 s of data}&\multicolumn{2}{|c|}{Input: 5 s of data}&\multicolumn{2}{|c}{Input: 10 s of data}\\
\hline
    instrument &proposed[\%]&baseline[\%]&proposed[\%]&baseline[\%]&proposed[\%]&baseline[\%]\\
 \hline
 drums&84.73&94.00&86.84&95.24&88.91&94.62\\
 bass&51.01&51.78&59.20&61.54&64.87&70.32\\
 piano&77.21&39.40&82.00&43.27&84.30&45.30\\
 guitar&76.50&-&80.18&-&82.70&-\\
 others&82.84&-&83.34&-&82.20&-\\
 \hline
 \end{tabular}
 }
 \caption{kNN-based classification accuracy using each embedding representation. The lines “proposed” and “baseline” show the results using the proposed method and using the separated signals as input to the individual networks of the conventional method~\cite{has2022}. }
 \label{table:knn}
\end{table}

\clearpage

\clearpage
 \begin{table}[th]
\setlength{\tabcolsep}{4pt}
\centering
\scalebox{0.8}{
\begin{tabular}{ccccccc||c|c|c|c|c}
\hline
\multicolumn{7}{c||}{Method}&\multicolumn{5}{|c}{Instrument[\%]}\\
\hline
 &baseline&norm&psd&basic&add&pre&drums&bass&piano&guitar&others\\
 \hline
 (a)&$\checkmark$&&&&&&90.19&40.00&41.28&-&-\\
 (b)&&&&&&$\checkmark$&84.31&34.65&33.72&41.46&71.06\\
 (c)&&&&$\checkmark$\textreferencemark&&&83.54&17.88&24.87&29.12&26.14\\
 (d)&&$\checkmark$&&$\checkmark$\textreferencemark&&&88.99&19.54&26.99&32.18&31.63\\
(e)&&$\checkmark$&&$\checkmark$\textreferencemark&&$\checkmark$&90.19&14.66&26.73&31.26&42.55\\
 (f)&&$\checkmark$&$\checkmark$&$\checkmark$&&&92.64&55.69&56.09&38.00&65.95\\
(g)&&$\checkmark$&$\checkmark$&$\checkmark$&$\checkmark$&&92.43&60.47&64.04&54.27&76.28\\
(h)&&$\checkmark$&$\checkmark$&$\checkmark$&&$\checkmark$&95.91&63.08&58.97&52.25&80.74\\
(i)&&$\checkmark$&$\checkmark$&$\checkmark$&$\checkmark$&$\checkmark$&95.80&61.91&67.37&57.32&80.69\\
 \hline
 \end{tabular}
 }
 \caption{kNN-based classification accuracy using the pseudo piece. The baseline is our conventional method~\cite{has2022}. The “norm,” “psd,” “basic,” “add,” and “pre” mean using the norm loss, the pseudo musical pieces, the basic triplets, the additional triplets, and the pre-training, respectively. \textreferencemark When learning without the pseudo musical pieces, the basic triplet sampling method is shown in Figure~\ref{nopsdtrip}, and the additional triplet cannot be created. Only the results of the five-second segment are shown.}
 \label{table:knnsub}
 \end{table}
\subsubsection{Capability to represent separated embeddings}
Table~\ref{table:knnsub} shows the evaluation results for each subspace using pseudo musical pieces. 
We can see that our proposed method, including pre-training, creating the pseudo musical pieces, and training with the additional triplets, is effective, and using a combination of these methods can lead to higher scores than conventional methods.
\par
The result (b) shows that the pre-trained model performed better than random prediction but worse than the conventional method.
The results (c)--(e) show that the models trained without the pseudo musical pieces (Figure~\ref{nopsdtrip}) did not work well even when used with the norm loss and the pre-training. Considering that only the drum score is high, these results suggest that all subspaces represent the drum's feature, which is consistent with the concern raised in Section~\ref{pseudo} that all subspaces might be learned on the same criteria.
We can see that the models trained with the pseudo musical pieces work well as shown in the results (f)--(i), and the additional triplet can improve the accuracy as shown by comparing (f) and (g); the pre-training also can improve the accuracy as shown by comparing (f) and (h), especially in low-accuracy instruments. These results mean that these methods can help separation for each subspace.
\par
\begin{figure}[t]
    \begin{minipage}[t]{\columnwidth}
        \centering
        \includegraphics[scale=0.3]{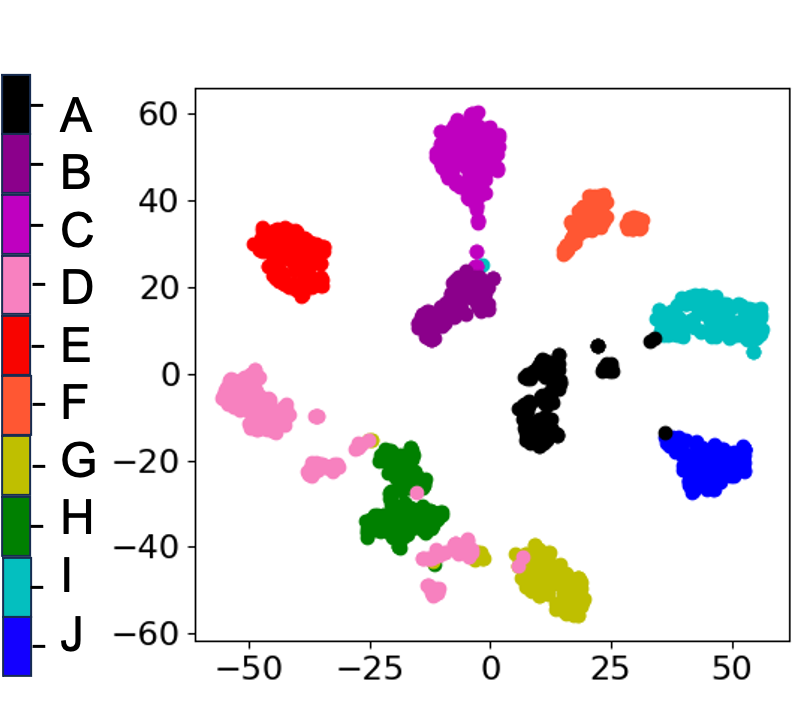}
        \subcaption{Drums subspace}
    \end{minipage}
    \\
    \begin{minipage}[t]{0.45\columnwidth}
        \centering
        \includegraphics[scale=0.3]{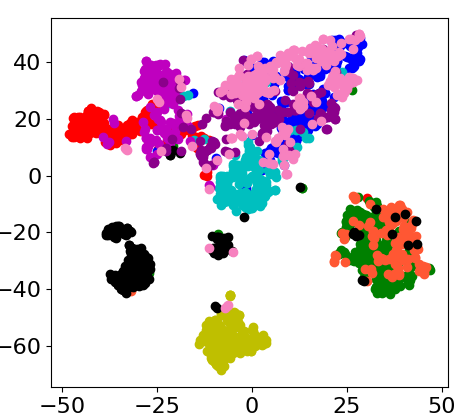}
        \subcaption{Bass subspace}
    \end{minipage}
    \hfill
    \begin{minipage}[t]{0.45\columnwidth}
        \centering
        \includegraphics[scale=0.3]{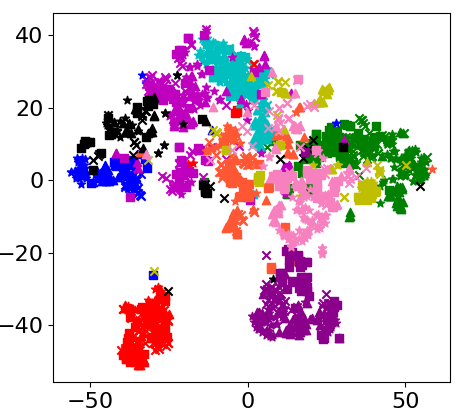}
        \subcaption{Piano subspace}
    \end{minipage}
    \hfill
    \begin{minipage}[t]{0.45\columnwidth}
        \centering
        \includegraphics[scale=0.3]{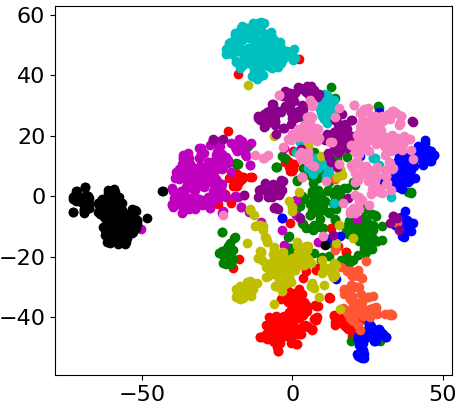}
        \subcaption{Guitar subspace}
    \end{minipage}
    \hfill
    \begin{minipage}[t]{0.45\columnwidth}
        \centering
        \includegraphics[scale=0.3]{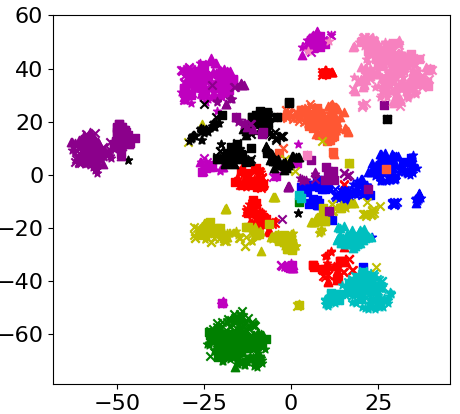}
        \subcaption{Others subspace}
    \end{minipage}

\caption{Examples of visualized embedding representations extracted from five-second segments in the test pseudo musical pieces. In each instrument, the same color represents the same instrument label. A set of segments plotted in the same color for one instrument includes segments with different labels for other instruments. For example in (a), a segment with label $\rm{(A, B)^{(dr, else)}}$ and a segment with label $\rm{(A, C)^{(dr, else)}}$ are plotted with the same color as they have the same drum label. This also includes segments with exactly the same label, such as $\rm{(A, B)^{(dr, else)}}$ and $\rm{(A, B)^{(dr, else)}}$, which are different time frames from the same musical piece. We can see in Figure~\ref{emb40-2} the visualization when different colors are assigned to segments that have the same instrument label but different labels for other instruments.}
\label{emb40}
\end{figure}
\begin{figure}[t]
    \begin{minipage}[t]{\columnwidth}
        \centering
        \includegraphics[scale=0.35]{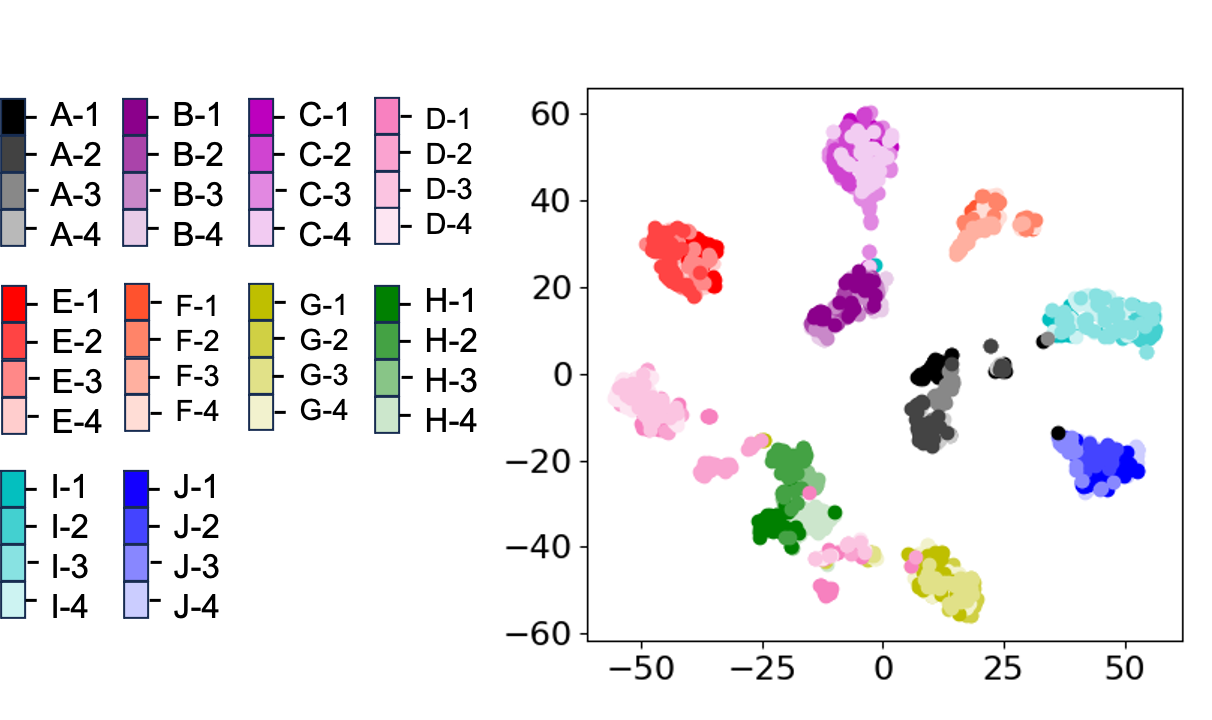}
        \subcaption{Drums subspace}
    \end{minipage}
    \\
    \begin{minipage}[t]{0.45\columnwidth}
        \centering
        \includegraphics[scale=0.35]{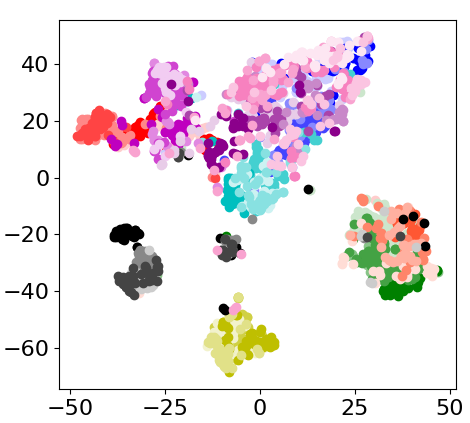}
        \subcaption{Bass subspace}
    \end{minipage}
    \hfill
    \begin{minipage}[t]{0.45\columnwidth}
        \centering
        \includegraphics[scale=0.35]{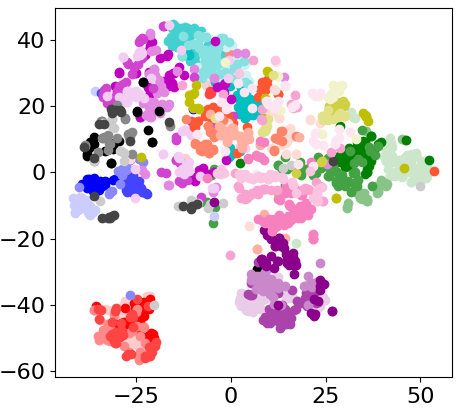}
        \subcaption{Piano subspace}
    \end{minipage}
    \hfill
    \begin{minipage}[t]{0.45\columnwidth}
        \centering
        \includegraphics[scale=0.35]{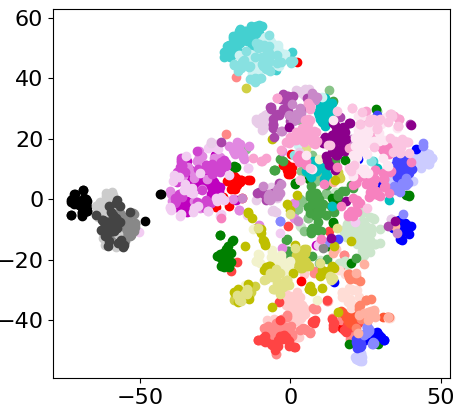}
        \subcaption{Guitar subspace}
    \end{minipage}
    \hfill
    \begin{minipage}[t]{0.45\columnwidth}
        \centering
        \includegraphics[scale=0.35]{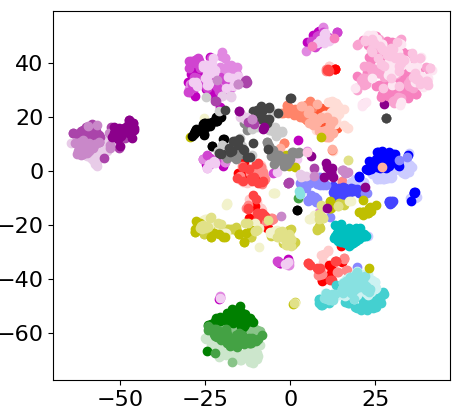}
        \subcaption{Others subspace}
    \end{minipage}
\caption{These are the same results as in Figure~\ref{emb40} but the color coding is different, where only segments with the same label for all instruments are plotted in the same color. For example in (a), a segment with label $\rm{(A, B)^{(dr, else)}}$ and a segment with label $\rm{(A, C)^{(dr, else)}}$ have the same drum label but are plotted with different colors. The color map in this figure denotes such labels with A-1, A-2, etc. When evaluating by kNN using the pseudo musical pieces in Section~\ref{pseknn}, the same color in Figure~\ref{emb40} was regarded as the same label, but segments with the same color in this figure were not used for reference.}
\label{emb40-2}
\end{figure}

All five-second segments divided from the 40 pseudo musical pieces used in the test are plotted and visualized in two dimensions in Figure~\ref{emb40} and Figure~\ref{emb40-2}. It can be seen that the pieces with the same instrumental sound labels are close to each other.

\subsubsection{Instrumental sound identification accuracy}
\begin{figure}[h]
    \begin{minipage}[t]{0.48\columnwidth}
        \centering
        \includegraphics[scale=0.25]{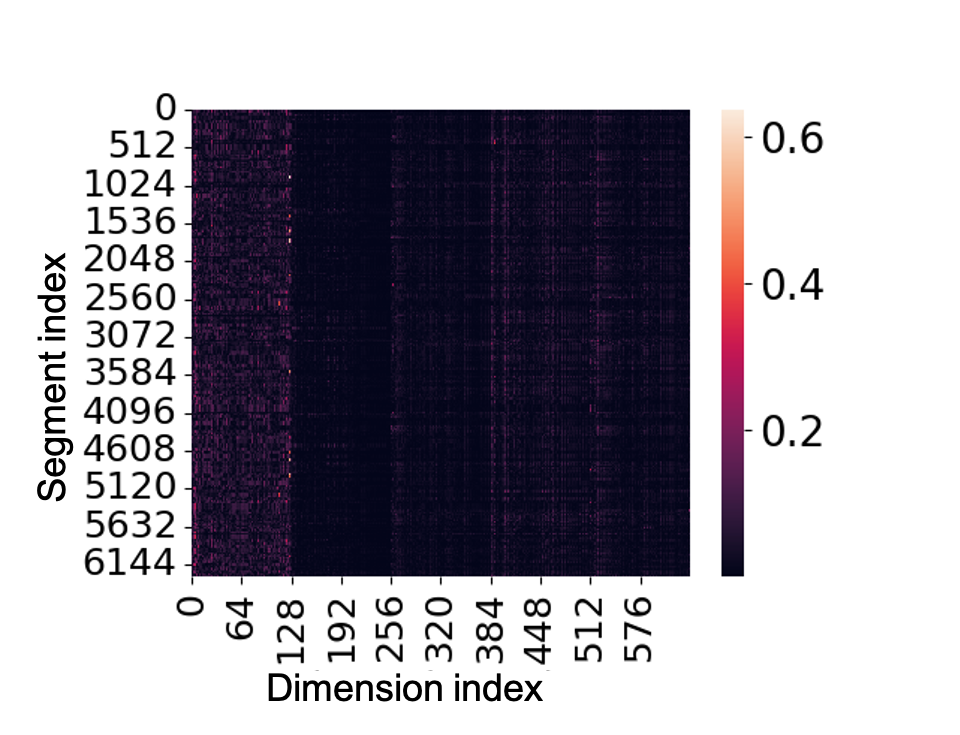}
        \subcaption{Drums subspace}
    \end{minipage}
    \begin{minipage}[t]{0.48\columnwidth}
        \centering
        \includegraphics[scale=0.25]{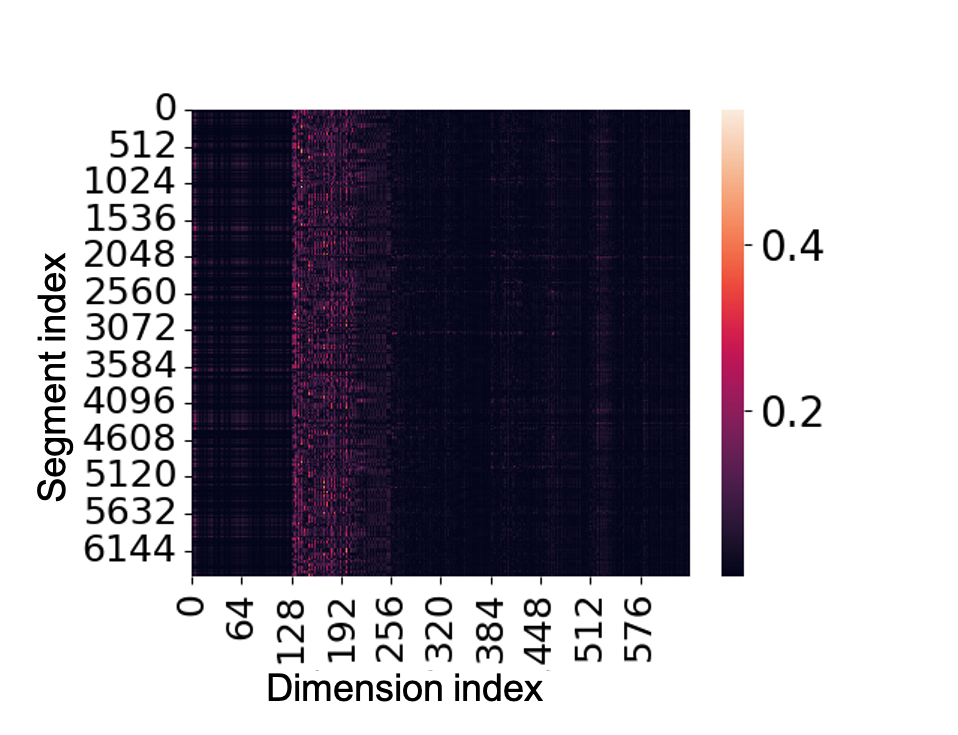}
        \subcaption{Bass subspace}
    \end{minipage}
    \begin{minipage}[t]{0.48\columnwidth}
        \centering
        \includegraphics[scale=0.25]{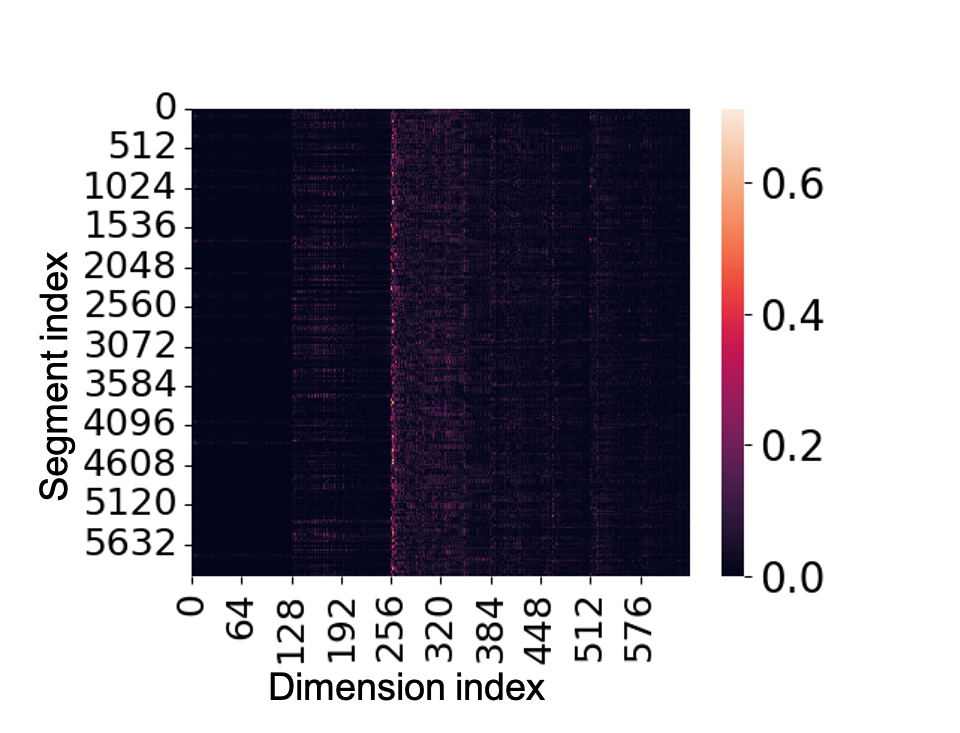}
        \subcaption{Piano subspace}
    \end{minipage}
    \begin{minipage}[t]{0.48\linewidth}
        \centering
        \includegraphics[scale=0.25]{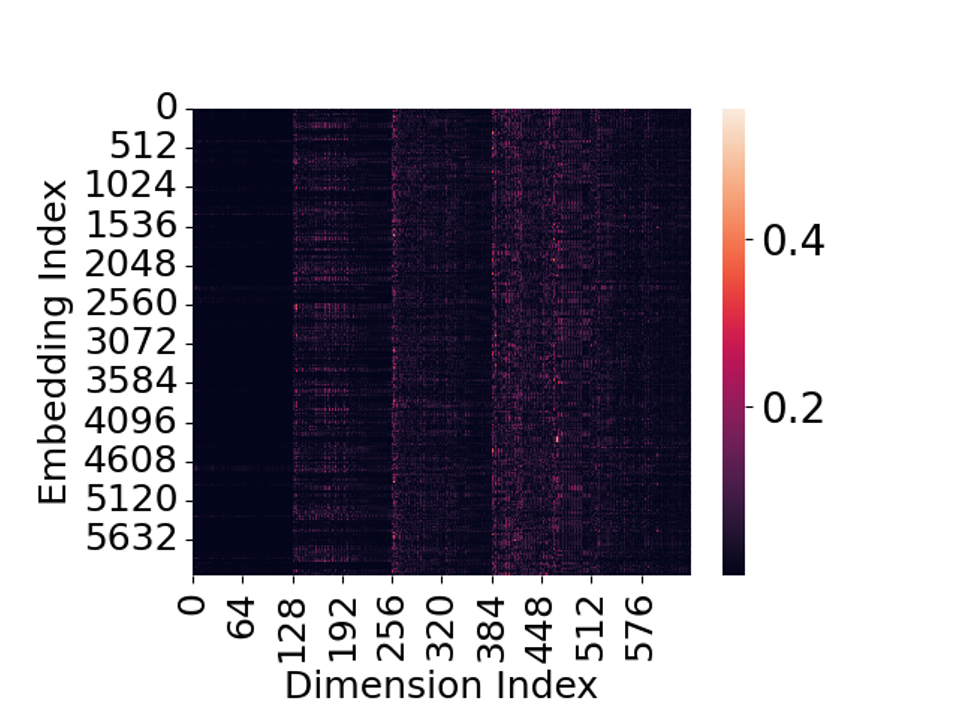}
        \subcaption{Guitar subspace}
    \end{minipage}
    \begin{minipage}[t]{0.48\columnwidth}
        \centering
        \includegraphics[scale=0.25]{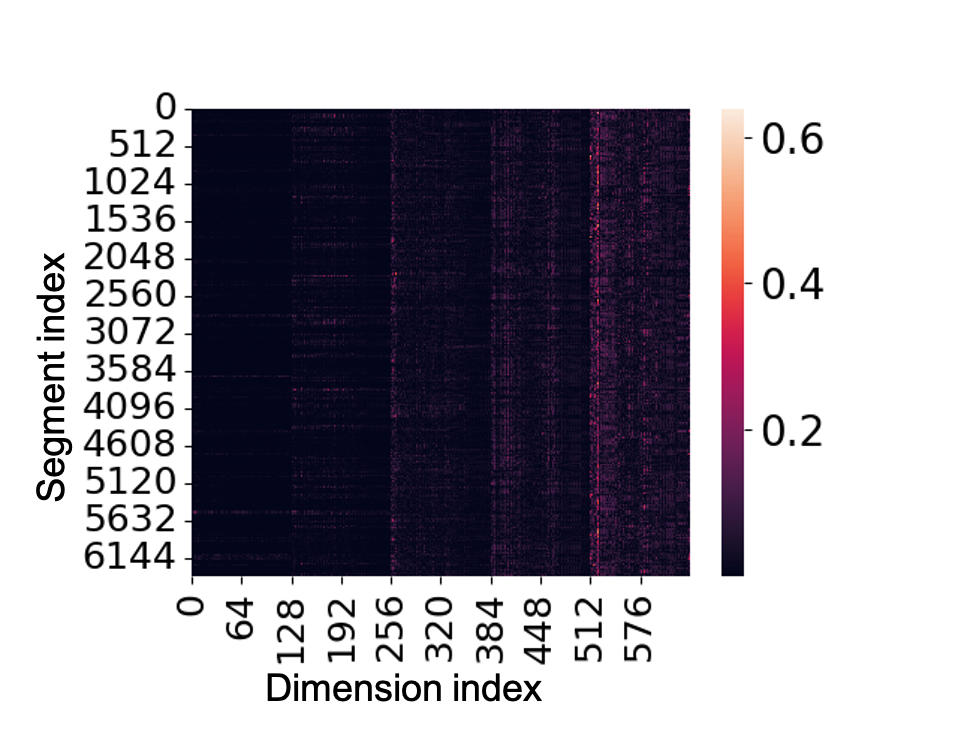}
        \subcaption{Others subspace}
    \end{minipage}
\caption{For each instrument, absolute values are taken for the output embeddings (without masking) when inputting five-second segments of the individual instrument sounds in the test set, and stacked vertically. The vertical axis is the index of segments, and the horizontal axis is the index of dimensions.}
\label{feat}
\end{figure}

Table~\ref{table:recg} shows the results of instrumental sound identification accuracy. The pre-training is effective for making the unsounded instruments' subspaces zero vector. Using the norm loss can also improve the accuracy. We can see that only the corresponding subspace retains the values when inputting the individual instrumental sound as in Figure~\ref{feat}.

\begin{table}[h]
\centering
\begin{center}
\scalebox{0.85}{
    \begin{tabular}{cccccc}
        \hline
        &drums (\%)&bass (\%)&piano (\%)&guitar (\%)&others (\%)\\
        \hline
        wo/pre, wo/norm&40.27&0.39&67.00&10.10&10.75\\
        w/pre, wo/ norm&67.94&76.89&78.65&67.01&95.66\\
        w/pre, w/ norm&84.48&96.04&90.27&76.85&92.24\\
        \hline
    \end{tabular}
}
\end{center}
\caption{Instrumental sound recognition rate when each individual instrument sound is input. “wo/pre” means using pre-training, and “wo/norm” means using norm loss. Each line name can be rephrased corresponding to the results in Table~\ref{table:knnsub} as follows:
“wo/pre, wo/norm,” to psd+basic+add,
“w/pre, wo/norm,” to psd+basic+add+pre,
and “w/pre, w/norm,” to norm+psd+basic+add+pre.}
\label{table:recg}
\end{table}

\clearpage


\subsubsection{Correlation between distance matrices}
\begin{figure}[h]
\begin{tabular}{cc}
  \begin{minipage}[t]{0.4\hsize}
 \centering
 \includegraphics[width=1.1\columnwidth]
 {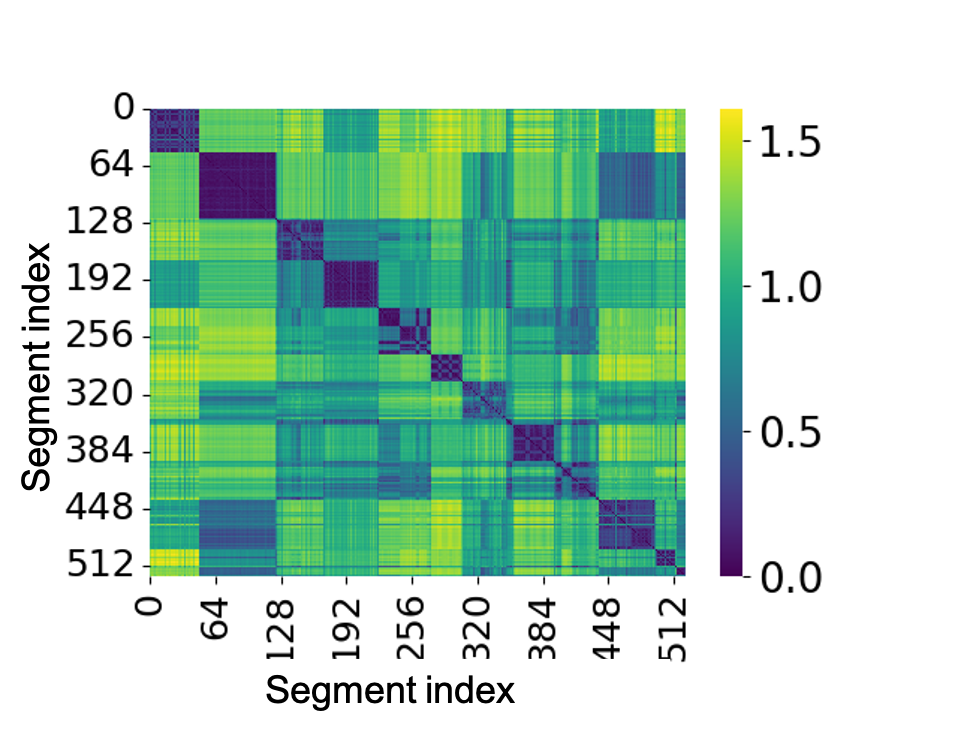}
 \caption*{(a) Using the drums subspace representation with drums sound input}
  \end{minipage}&
  \hspace{-8pt}
  \begin{minipage}[t]{0.4\hsize}
 \centering
 \includegraphics[width=1.1\columnwidth]{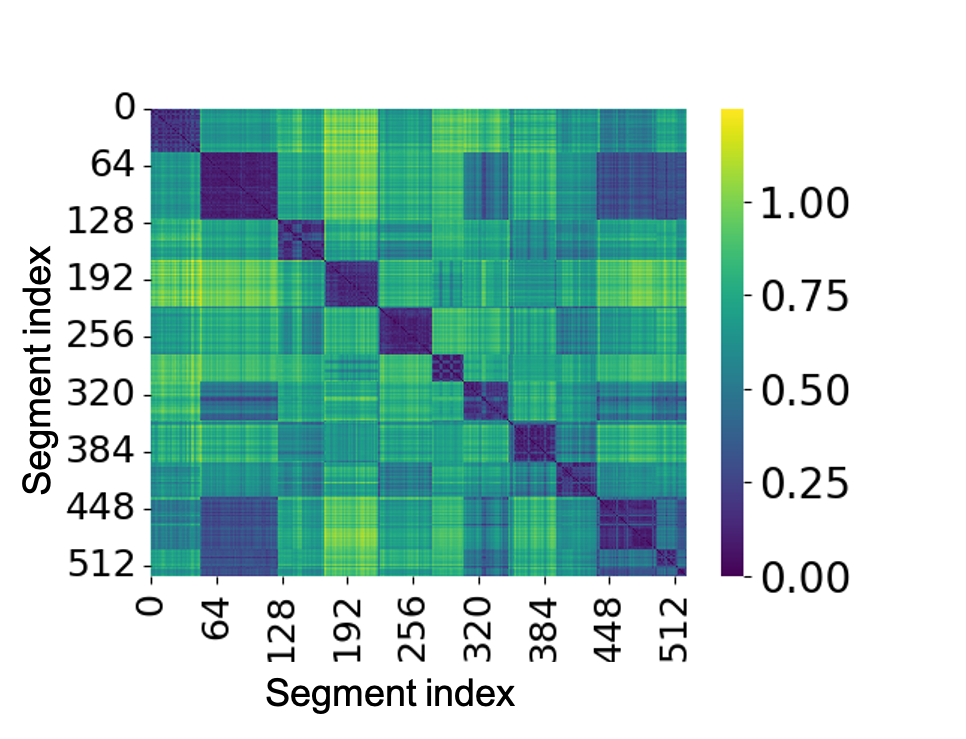}
 \caption*{(b) Using the drums subspace representation with mixed sound input}
 \end{minipage}
 \end{tabular}
 \caption{Distance matrices between the representations of segments from musical pieces in the test set. Here, we show an example of 10 musical pieces. This is the result of the model trained using the pseudo musical pieces; “w/psd”.}
 \label{fig:dist}
\end{figure}

Table~\ref{table:corr} shows the correlation between the distance matrix when an individual instrument sound is input and the distance matrix when a mixed sound containing that instrument sound is input and masked. It can be seen that the correlation is higher when pseudo musical pieces were used, which suggests that using pseudo musical pieces can help each subspace represent the target instrumental feature. A visualization of the two distance matrices is shown in Figure~\ref{fig:dist} taking the drums as an example. We can see a similar pattern in the similarity between the musical pieces both when the instrumental signals are input and when the mixed signals containing them are input and masked.
\par
Table~\ref{table:corr_mix} shows the correlation between pairs of distance matrices across the subspaces. The results with the individual network with clean individual instrumental sound input, shown for reference, show low correlations, indicating that the correlation of embedding for each instrument should inherently be low. However, without the pseudo musical pieces, the correlation between subspaces is high, indicating that all spaces are learned with the same criteria. The correlation for the proposed method using the pseudo musical pieces is low, which suggests that the proposed method using the pseudo musical pieces can learn the instrument-dependent features in each subspace.

\begin{table}[h]
\centering
\begin{center}
\scalebox{0.8}{
    \begin{tabular}{cccccc}
        \hline
        &drums&	bass&	piano&	guitar&	others\\
        \hline
        wo/ psd&0.6418&0.2357&0.1651&0.3777&0.2743\\
        w/ psd&0.6159&0.4782&0.3608&0.4115&0.3243\\
        \hline
    \end{tabular}
}
\end{center}
\caption{Correlation between distance matrices in the same subspace with individual instrumental sound input and with mixed sound input. 
Each line name can be rephrased corresponding to the results in Table~\ref{table:knnsub} as follows:
“wo/psd” to norm+basic+pre,
and “w/ psd” to norm+psd+basic+add+pre.}
\label{table:corr}
\end{table}

\begin{table}[h]
\centering
\begin{center}
\scalebox{0.8}{
    \begin{tabular}{cccccc}
        \hline
        &\multicolumn{5}{c}{w/ individual clean sound (reference)}\\
        \hline
        &drums&	bass&	piano&	guitar&	others\\
        \hline
        drums&-&0.000112&0.0133&0.0509&0.0123\\
        bass&-&-&0.0320&0.0567&0.0531\\
        piano&-&-&-&0.135&0.209\\
        guitar&-&-&-&-&0.156\\
        others&-&-&-&-&-\\	
        \hline
        &\multicolumn{5}{c}{wo/ psd}\\
        \hline
        &drums&	bass&	piano&	guitar&	others\\
        \hline
        drums&-&0.610&0.359&0.224&0.226\\
        bass&-&-&0.444&0.323&0.434\\
        piano&-&-&-&0.304&0.460\\
        guitar&-&-&-&-&0.246\\
        others&-&-&-&-&-\\	
        \hline
        &\multicolumn{5}{c}{w/ psd}\\
        \hline
        &drums&	bass&	piano&	guitar&	others\\
        \hline
        drums&-&-0.0298&-0.100&-0.00900&-0.0929\\
        bass&-&-&-0.0488&0.0309&0.0315\\
        piano&-&-&-&-0.170&0.0697\\
        guitar&-&-&-&-&0.00354\\
        others&-&-&-&-&-\\			
        \hline
    \end{tabular}
    }
\end{center}
\caption{Correlation between pairs of distance matrices across the subspaces. The “wo/psd” and “w/psd” represent the same mean as Table~\ref{table:corr}.}
\label{table:corr_mix}
\end{table}

\subsubsection{Subjective evaluation}
The matching rate with the answers on \textit{overall} was high for \textit{rhythm} and \textit{melody}, but low for \textit{timbre}. In other words, different labels were assigned to the same sample set when focusing on \textit{timbre} compared to focusing on \textit{overall}. Therefore, we evaluated the model using these two types of answers.
\par
The results are shown in Table~\ref{table:subTR}, and the number of answers for each instrument on the two perspectives are shown in Table~\ref{tab:eval_prev_num}. 
We can see from Table~\ref{tab:eval_prev_num} that there is more agreement among subjects in their answers in test 2 than in test 1. Moreover, there are fewer answers for \textit{timbre} than for \textit{overall} because the subjects cannot select N/A for \textit{overall}, but they can for \textit{timbre}, and here, N/A (after taking agreement) is omitted from the evaluation. 
\par
As shown in “set 2” in Table~\ref{table:subTR}, different time segments within the same piece are perceived by humans as similar to each other, and their distances are also small in the distance metric learned by the proposed method.
\par
Compared with the baseline~\cite{has2022}, the matching rates of the proposed method for the drums and bass are comparable, and that for the piano is better.
\par
Although test 1 is less accurate than test 2, accuracy is improved in drums, piano, and guitar in the evaluation using answers focusing on timbre compared with using answers focusing on overall. This suggests that the model is trained to represent similarity mainly focusing on timbre. 
There are two possible reasons for this. First, since the model is trained using different temporal segments of instrumental sounds from the same musical piece as positive examples, it may learn to minimize the distance between segments with different melodies. This suggests that the model potentially learns to ignore melodic information. Second, the embeddings of relatively short segments, i.e., three seconds, are temporally averaged, which likely discards the sequential structure of the melody. Additionally, rhythmic information may also be lost during this process. As a result, the model tends to capture timbral characteristics that remain consistent in the time direction of features.
We consider that if we can design a model that captures the structure of the time direction so that melody and rhythm can be considered, it will be possible to obtain a music similarity that is also compatible with human perception when focusing on the overall similarity.


\begin{table}[ht]
\centering
\begin{center}
\scalebox{0.95}{
    \begin{tabular}{c|ccccc}
    \hline
    &\multicolumn{5}{|c}{Evaluation with answers on \textbf{overall} (\%)}\\
          \hline
    	&drums&	bass&	piano&	guitar&	others\\
     \hline
     test 1\\
     \hline
     baseline	&56.5$\pm$3.3&56.0$\pm$3.2&58.4$\pm$3.4	&-	&-\\
    proposed	&56.7$\pm$3.3 &68.3$\pm$3.1&57.7$\pm$3.4&61.7$\pm$3.2&60.0$\pm$3.1\\
    \hline
    test 2\\
    \hline
    baseline	&95.5$\pm$1.0&83.8$\pm$1.8&74.1$\pm$1.9&-	&-\\
    proposed	&95.7$\pm$1.0&89.1$\pm$1.5&88.7$\pm$1.4&93.0$\pm$1.2&92.5$\pm$1.3\\
    \hline
    \hline
     &\multicolumn{5}{|c}{Evaluation with answers on \textbf{timbre} (\%)}\\
     \hline
    	&drums&	bass&	piano&	guitar&	others\\ 
        \hline
     test1\\
     \hline
     baseline	&61.6$\pm$4.6&55.0$\pm$4.9&59.1$\pm$5.5&-	&-\\
    proposed	&66.1$\pm$4.5&69.3$\pm$4.7&70.6$\pm$5.2&73.8$\pm$4.5&64.7$\pm$4.8\\
    \hline
    test 2\\
    \hline
    baseline &	96.2$\pm$1.0&81.7$\pm$1.9&74.5$\pm$1.9&-&-\\
    proposed	&96.5$\pm$0.9&87.4$\pm$1.6&89.4$\pm$1.4&93.7$\pm$1.2&93.6$\pm$1.2\\
    \hline
    \end{tabular}
    }
\end{center}
\caption{Matching rate between the model's results and subjects' results focusing on overall and timbre, respectively. The baseline is our previous method~\cite{has2022}.}
\label{table:subTR}
\end{table}

\begin{table}[t]
\centering
\caption{Number of answers using results with 80\% agreement among subjects focusing on overall and timbre, respectively.}
\vspace{-8pt}
\begin{center}
\begin{tabular}{c|ccccc}
      \hline
      &\multicolumn{5}{|c}{Evaluation with answers on \textbf{overall}}\\
  \hline
	&drums&	bass&	piano&	guitar&	others\\
 \hline
test 1	&912&949&836&925&977\\
test 2	&2143&1792&2082&2096&1926\\
\hline
      \hline
      &\multicolumn{5}{|c}{Evaluation with answers on \textbf{timbre}}\\
  \hline

	&drums	&bass	&piano	&guitar	&others\\
 \hline
test 1	&463&420&330&412&414\\
test 2	&1926&1745&2044&1993&1892\\
\hline
    \end{tabular}
\end{center}
\vspace{-16pt}
\label{tab:eval_prev_num}
\end{table}

\section{Conclusions}

In this paper, we proposed a method of computing similarities focusing on each instrumental sound using mixed signals as input in one network, which extracts a single similarity embedding space with separated dimensions for each instrument using CSNs. To successfully train the network, we implemented new ideas for the training, such as the use of pseudo musical pieces, a norm loss, and pre-training. Experimental results showed the effectiveness of our strategies and that the selection of similar musical pieces focusing on each instrumental sound by the proposed method can obtain human acceptance, especially when focusing on timbre. Future work is to design a model that captures the structure of the time direction so that melody and rhythm can be considered.

\section*{Acknowledgment}
This work was partly supported by JST CREST Grant Number JPMJCR19A3 and Grant-in-Aid for JSPS Fellows JP24KJ1253, Japan.

\clearpage
\section*{References}
\printbibliography

@inproceedings{has2022,
    author = {Y. Hashizume and L. Li and T. Toda},
    year = {2022},
    pages={33--38},
    title = {Music similarity calculation of individual instrumental sounds using metric learning},
    booktitle={Asia-Pacific Signal and Information Processing Association Annual Summit and Conference}
}

@INPROCEEDINGS{veit2017,
  author={Veit, Andreas and Belongie, Serge and Karaletsos, Theofanis},
  booktitle={IEEE Conference on Computer Vision and Pattern Recognition}, 
  title={Conditional similarity networks}, 
  year={2017},
  pages={1781-1789},
}

@inproceedings{song2012,
    author = {Song, Yading and Dixon, Simon and Pearce, Marcus},
    year = {2012},
    pages={395--410},
    title = {A survey of music recommendation systems and future perspectives},
    booktitle={International Symposium on Computer Music Modelling and Retrieval},
}

@ARTICLE{casey2008,
    author={Casey, Michael A. and Veltkamp, Remco and Goto, Masataka and Leman, Marc and Rhodes, Christophe and Slaney, Malcolm},
    journal={Proceedings of the IEEE},
    title={Content-based music information retrieval: Current directions and future challenges},
    year={2008},
    volume={96},
    number={4},
    pages={668--696},
}

@inproceedings{hamel2010,
    title={Learning features from music audio with deep belief networks},
    author={Philippe Hamel and Douglas Eck},
    booktitle={International Society for Music Information Retrieval Conference},
    year={2010},
    pages={339--344}
}

@article{elbir2020,
    author = {Elbir, A. and Aydin, N.},
    title = {Music genre classification and music recommendation by using deep learning},
    journal = {Electronics Letters},
    volume = {56},
    number = {12},
    year={2020},
    pages = {627--629},
}

@article{fathollahi2021,
    title={Music similarity measurement and recommendation system using convolutional neural networks},
    author={Fathollahi, Mohamadreza Sheikh and Razzazi, Farbod},
    journal={International Journal of Multimedia Information Retrieval},
    volume={1},
    number={10},
    pages={43--53},
    year={2021}
}

@INPROCEEDINGS{lu2017,
    author={Lu, Rui and Wu, Kailun and Duan, Zhiyao and Zhang, Changshui},
    title={Deep ranking: Triplet matchNet for music metric learning},
    year={2017},
    volume={},
    number={},
    pages={121--125},
    booktitle={IEEE International Conference on Acoustics, Speech and Signal Processing},
}

@misc{clevelan2020,
    author = {Cleveland, Joseph and Cheng, Derek and Zhou, Michael and Joachims, Thorsten and Turnbull, Douglas},
    title = {Content-based music similarity with triplet networks},
    publisher = {arXiv},
    year = {2020},
    note={arXiv:2008.04938}
}

@INPROCEEDINGS{lee2020,
    author={Lee, Jongpil and Bryan, Nicholas J. and Salamon, Justin and Jin, Zeyu and Nam, Juhan},
    title={Disentangled multidimensional metric learning for music similarity},
    year={2020},
    volume={},
    number={},
    pages={6--10},
    booktitle={IEEE International Conference on Acoustics, Speech and Signal Processing},
}

@inproceedings{choi2019,
    title={Zero-shot learning for audio-based music classification and tagging},
    author={Jeong Choi and Jongpil Lee and Jiyoung Park and Juhan Nam},
    booktitle={International Society for Music Information Retrieval Conference},
    pages={67--74},
    year={2019},
}

@InProceedings{ailon2015,
    author={Hoffer, Elad and Ailon, Nir},
    title={Deep metric learning using triplet network},
    booktitle={Similarity-Based Pattern Recognition},
    year={2015},
    pages={84--92}
}

@InProceedings{bengio2013,
    author={Y. Bengio},
    title={Deep learning of representations: Looking forward},
    booktitle={International Conference on Statistical Language and Speech Processing},
    year={2013},
    pages={1--37}
}

@InProceedings{hsu2018,
      title={Hierarchical generative modeling for controllable speech synthesis}, 
      author={Wei-Ning Hsu and Yu Zhang and Ron J. Weiss and Heiga Zen and Yonghui Wu and Yuxuan Wang and Yuan Cao and Ye Jia and Zhifeng Chen and Jonathan Shen and Patrick Nguyen and Ruoming Pang},
      year={2019},
      booktitle={International Conference on Learning Representations},
      page={1--27}
}

@INPROCEEDINGS{hsu2019,
  author={Hsu, Wei-Ning and Zhang, Yu and Weiss, Ron J. and Chung, Yu-An and Wang, Yuxuan and Wu, Yonghui and Glass, James},
  booktitle={IEEE International Conference on Acoustics, Speech and Signal Processing}, 
  title={Disentangling correlated speaker and noise for speech synthesis via data augmentation and adversarial factorization}, 
  year={2019},
  pages={5901--5905},
}

@InProceedings{luo2019,
    author={Y-J. Luo, K. Agres and D. Herremans},
    title={Learning disentangled representations of timbre and pitch for
musical instrument sounds using gaussian mixture variational autoencoders},
    booktitle={International Society for Music Information Retrieval Conference},
    year={2019},
    pages={746--753}
}

@INPROCEEDINGS{tanaka21,
  author={Tanaka, Keitaro and Nishikimi, Ryo and Bando, Yoshiaki and Yoshii, Kazuyoshi and Morishima, Shigeo},
  booktitle={IEEE International Conference on Acoustics, Speech and Signal Processing}, 
  title={Pitch-timbre disentanglement Of musical instrument sounds based on VAE-based metric learning}, 
  year={2021},
  pages={111-115},
}

@inproceedings{manilow2019,
    author = {Manilow, Ethan and Wichern, Gordon and Seetharaman, Prem and Le Roux, Jonathan},
    year = {2019},
    pages = {45--49},
    title = {Cutting music source separation some Slakh: A dataset to study the impact of training data quality and quantity},
    booktitle={IEEE Workshop on Applications of Signal Processing to Audio and Acoustics},
}

@article{hennequin2020,
    author = {Hennequin, Romain and Khlif, Anis and Voituret, Felix and Moussallam, Manuel},
    year = {2020},
    pages = {2154},
    title = {Spleeter: A fast and efficient music source separation tool with pre-trained models},
    volume = {5},
    journal = {Journal of Open Source Software},
}

@article{laurens2008,
    author  = {Laurens van der Maaten and Geoffrey Hinton},
    title   = {Visualizing data using t-SNE},
    journal = {Journal of Machine Learning Research},
    year    = {2008},
    volume  = {9},
    number  = {86},
    pages   = {2579--2605},
}

@inproceedings{jansson2017,
    title={Singing voice separation with deep U-Net convolutional networks},
    author={Andreas Jansson and Eric J. Humphrey and Nicola Montecchio and Rachel M. Bittner and Aparna Kumar and Tillman Weyde},
    booktitle={International Society for Music Information Retrieval Conference},
    year={2017}
}

@misc{ifpi,
    author  = {IFPI},
    title   = {Global music report 2024},
    note    = {\url{https://www.ifpi.org/wp-content/uploads/2024/04/GMR_2024_State_of_the_Industry.pdf}},
    year    = {2024}
}

@misc{apple,
    author  = {{Apple Inc}},
    title   = {Apple music},
    note    = {\url{https://www.apple.com/jp/apple-music/}},
    year    = {2024}
}

@inproceedings{logan2004,
  title={Music recommendation from song sets},
  author={Beth Logan},
  booktitle={International Society for Music Information Retrieval Conference},
  pages={4 pages},
  year={2004}
}

@article{goldberg1992,
    author = {Goldberg, David and Nichols, David and Oki, Brian M. and Terry, Douglas},
    title = {Using collaborative filtering to weave an information tapestry},
    year = {1992},
    volume = {35},
    number = {12},
    pages = {61--70},
    journal = {Communications of the ACM},
}

@inproceedings{dror2011,
    author = {Koenigstein, Noam and Dror, Gideon and Koren, Yehuda},
    year = {2011},
    pages = {165--172},
    title = {Yahoo! music recommendations: Modeling music ratings with temporal dynamics and item taxonomy},
    booktitle = {ACM Recommender Systems},
}

@article{chen2005,
    author = {Chen, Hung-Chen and Chen, Arbee},
    year = {2005},
    pages = {113--132},
    title = {A music recommendation system based on music and user grouping},
    volume = {24},
    journal = {Journal of Intelligent Information Systems},
}

@INPROCEEDINGS{logan2001,
    author={Logan, B. and Salomon, A.},
    title={A music similarity function based on signal analysis},
    year={2001},
    volume={},
    number={},
    pages={745--748},
    booktitle={IEEE International Conference on Multimedia and Expo},
}

@inproceedings{wolff2015,
    author = {Wolff, Daniel and Macfarlane, Andrew and Weyde, Tillman},
    year = {2015},
    pages = {24--30},
    title = {Comparative music similarity modelling using transfer learning across user groups},
    booktitle={International Society for Music Information Retrieval Conference},
}

@inproceedings{park2018,
    author    = {Jiyoung Park and Jongpil Lee and
    Jangyeon Park and Jung-Woo Ha and Juhan Nam},
    title     = {Representation learning of music using artist labels},
    pages     = {717--724},
    year      = {2018},
    booktitle = {International Society for Music Information Retrieval Conference},
}

@inproceedings{oord2013,
    author = {van den Oord, Aaron and Dieleman, Sander and Schrauwen, Benjamin},
    booktitle = {Advances in neural information processing systems},
    pages = {1--9},
    title = {Deep content-based music recommendation},
    volume = {26},
    year = {2013},
}

@ARTICLE{tzane2001,
    author={Tzanetakis, G. and Cook, P.},
    journal={IEEE Transactions on Speech and Audio Processing},
    title={Musical genre classification of audio signals},
    year={2002},
    volume={10},
    number={5},
    pages={293-302},
}

@InProceedings{dror2012,
    title = 	 {The Yahoo! music dataset and KDD-Cup’11},
    author = 	 {Dror, Gideon and Koenigstein, Noam and Koren, Yehuda and Weimer, Markus},
    booktitle = 	 {Proceedings of KDD Cup 2011},
    pages = 	 {3--18},
    year = 	 {2012},
    volume = 	 {18},
    series = 	 {Proceedings of Machine Learning Research},
}

@ARTICLE{mcfee2012,
    author={McFee, Brian and Barrington, Luke and Lanckriet, Gert},
    journal={IEEE Transactions on Audio, Speech, and Language Processing},
    title={Learning content similarity for music recommendation},
    year={2012},
    volume={20},
    number={8},
    pages={2207--2218}
}

@InProceedings{ronneberger2015,
    author={Ronneberger, Olaf
    and Fischer, Philipp
    and Brox, Thomas},
    title={U-Net: Convolutional networks for biomedical image segmentation},
    booktitle={Medical Image Computing and Computer-Assisted Intervention},
    year={2015},
    pages={234--241},
}

@misc{cw,
title={CrowdWorks},
note= {https://crowdworks.jp}
}

@inproceedings{fujishima1999,
    title={Realtime chord recognition of musical sound: A system using common lisp music},
    author={Takuya Fujishima},
    booktitle={International Conference on Mathematics and Computing},
    year={1999},
}

@INPROCEEDINGS{whitman2002,
  author={Whitman, B. and Rifkin, R.},
  booktitle={IEEE Workshop on Multimedia Signal Processing}, 
  title={Musical query-by-description as a multiclass learning problem}, 
  year={2002},
  volume={},
  number={},
  pages={153-156}}

@article{gomez2006,
    author = {Gomez, Emilia},
    year = {2006},
    pages = {294--304},
    title = {Tonal description of polyphonic audio for music content processing},
    volume = {18},
    journal = {INFORMS Journal on Computing},
}

@inproceedings{hamel2013,
    title={Transfer learning in mir: Sharing learned latent representations for music audio classification and similarity},
    author={Philippe Hamel and Matthew E. P. Davies and Kazuyoshi Yoshii and Masataka Goto},
    booktitle={International Society for Music Information Retrieval Conference},
    year={2013},
}

@misc{li2015,
    title={Automatic instrument recognition in polyphonic music using convolutional neural networks},
    author={Peter Li and Jiyuan Qian and Tian Wang},
    year={2015},
    note={arXiv:1511.05520},
}

@inproceedings{hirvonen2015,
author = {Hirvonen, Toni},
year = {2015},
pages = {1--10},
title = {Classification of spatial audio location and content using convolutional neural networks},
volume = {2},
booktitle= {Audio Engineering Society Convention}
}

@inproceedings{choi2015,
  title={Auralisation of deep convolutional neural networks: Listening to learned features},
  author={Choi, Keunwoo and Fazekas, George and Sandler, Mark and Kim, Jeonghee},
  booktitle={International Society for Music Information Retrieval Conference},
  pages={26--30},
  year={2015}
}

@inproceedings{choi2016,
    title={Automatic tagging using deep convolutional neural networks},
    author={Choi, Keunwoo and Fazekas, Gyorgy and Sandler, Mark},
    booktitle={International Society for Music Information Retrieval Conference},
    pages={805—811},
    year={2016},
}

@inproceedings{lostanlen2016,
    title={Deep convolutional networks on the pitch spiral for music instrument recognition},
    author={Lostanlen, Vincent and Cella, Carmine-Emanuele},
    booktitle={International Society for Music Information Retrieval Conference},
    year={2016},
    pages={612--618}
}

@article{costa2017,
    author = {Costa, Yandre and Soares de Oliveira, Luiz and Silla, Carlos},
    year = {2017},
    pages = {1—37},
    title = {An evaluation of convolutional neural networks for music classification using spectrograms},
    volume = {52},
    journal = {Applied Soft Computing},
}

@inproceedings{wise2021,
  title={Attention augmented CNNs for musical instrument identification},
  author={Wise, Andrew and Maida, Anthony S and Kumar, Ashok},
  booktitle={European Signal Processing Conference},
  pages={376--380},
  year={2021}
}

@INPROCEEDINGS{pretet2020,
    author={Pretet, Laure and Richard, Gael and Peeters, Geoffroy},
    booktitle={IEEE International Conference on Acoustics, Speech and Signal Processing},
    title={Learning to rank music tracks using triplet loss},
    year={2020},
    volume={},
    number={},
    pages={511-515}
}

@article{pampalk2006,
    title={Computational models of music similarity and their application in music information retrieval},
    journal={PhD Thesis, Vienna University of Technology},
    pages={40 pages},
    author={Elias Pampalk},
    year={2006},
}

@inproceedings{aucouturier2002,
  title={Music similarity measures: What's the use?},
  author={Aucouturier, Jean-Julien and Pachet, Francois},
  booktitle={International Society for Music Information Retrieval Conference},
  volume={7},
  pages={339--340},
  year={2002}
}

@INPROCEEDINGS{casey2006,
    author={Casey, M. and Slaney, M.},
    booktitle={IEEE International Conference on Acoustics Speech and Signal Processing Proceedings},
    title={The importance of sequences in musical similarity},
    year={2006},
    volume={5},
    number={},
    pages={4 pages}
}

@inproceedings{schlter2013,
    title={Learning binary codes for efficient large-scale music similarity search},
    author={Jan Schluter},
    booktitle={International Society for Music Information Retrieval Conference},
    year={2013},
    page={6 pages}
}

@article{wolff2014,
    author = {Wolff Daniel, Weyde Tillman},
    year = {2014},
    pages = {109—136},
    title = {Learning music similarity from relative user ratings},
    volume = {17},
    journal = {Information Retrieval},
}

@inproceedings{cheng2020,
  title={Exploring acoustic similarity for novel music recommendation},
  author={Cheng, Derek and Joachims, Thorsten and Turnbull, Douglas R},
  booktitle={International Society for Music Information Retrieval Conference},
  pages={583--589},
  year={2020}
}

@inproceedings{gabbolini2021,
  title={An interpretable music similarity measure based on path interestingness},
  author={Gabbolini, Giovanni and Bridge, Derek},
  booktitle={International Society for Music Information Retrieval Conference},
  year={2021},
  pages={213--219},
}

@inproceedings{urbano2011,
  title={Audio music similarity and retrieval: Evaluation power and stability},
  author={Urbano, Julian and Martin, Diego and Marrero, Monica and Morato, Jorge},
  booktitle={International Society for Music Information Retrieval Conference},
  pages={597--602},
  year={2011},
}

@inproceedings{yoshii2006,
    author = {Yoshii, Kazuyoshi and Goto, Masataka and Komatani, Kazunori and Ogata, Tetsuya and Okuno, Hiroshi},
    year = {2006},
    pages = {296-301},
    title = {Hybrid collaborative and content-based music recommendation using probabilistic model with latent user preferences},
    booktitle = {International Conference on Music Information Retrieval}
}

@INPROCEEDINGS{iliadis2022,
    author={Iliadis, Lazaros Alexios and Sotiroudis, Sotirios P. and Kokkinidis, Kostas and Sarigiannidis, Panagiotis and Nikolaidis, Spiridon and Goudos, Sotirios K.},
    booktitle={International Conference on Modern Circuits and Systems Technologies},
    title={Music deep learning: A survey on deep learning methods for music processing},
    year={2022},
    volume={},
    number={},
    pages={1-4}
}

@misc{choi2018,
title={A tutorial on deep learning for music information retrieval},
author={Keunwoo Choi and György Fazekas and Kyunghyun Cho and Mark Sandler},
year={2018},
note={arXiv:1709.04396}
}

@inproceedings{senac2017,
    author = {Senac, Christine and Pellegrini, Thomas and Mouret, Florian and Pinquier, Julien},
    title = {Music feature maps with convolutional neural networks for music genre classification},
    year = {2017},
    booktitle = {International Workshop on Content-Based Multimedia Indexing},
    page={5 pages}
}

@misc{hung2018,
  title={Learning Disentangled Representations for Timber and Pitch in Music Audio},
  author={Yun-Ning Hung and Yian Chen and Yi-Hsuan Yang},
  year={2018},
  note={arXiv:1811.03271},
}

@INPROCEEDINGS{has25,
  author={Hashizume, Yuka and Toda, Tomoki},
  booktitle={2025 IEEE International Conference on Acoustics, Speech and Signal Processing}, 
  title={Investigation of perceptual music similarity focusing on each instrumental part}, 
  year={2025},
  volume={},
  number={},
  pages={1--5},}

\end{document}